\documentclass[aps,prd,preprintnumbers,nofootinbib,showpacs]{revtex4}%
\usepackage{graphicx}
\usepackage{bm,latexsym,amsmath,amssymb,amsfonts,mathrsfs}
\usepackage{color}
\usepackage{subfigure}
\input{colordvi.tex}
\usepackage{soul}
\usepackage{bm}

\begin{document}

\providecommand{\abs}[1]{\lvert#1\rvert}
\providecommand{\bd}[1]{\boldsymbol{#1}}

\preprint{KEK-TH-1999,HUPD-1711}
\vspace{2cm}
\title{Entanglement of the Vacuum between Left, Right, Future, and Past: \\
The Origin of Entanglement-Induced Quantum Radiation}
\vspace{1cm}

\author{Atsushi Higuchi}
\affiliation{Department of Mathematics, University of York, Heslington, York YO10 5DD, 
United Kingdom}

\author{Satoshi Iso}
\affiliation{KEK Theory Center, High Energy Accelerator Research Organization (KEK), 
Tsukuba, Ibaraki 305-0801, Japan}

\author{Kazushige Ueda}
\affiliation{Graduate School of Physical Sciences, Department of Physics,
Hiroshima University, Kagamiyama 1-3-1, Higashi-hiroshima, Hiroshima 739-8526, Japan}

\author{Kazuhiro Yamamoto}
\affiliation{Graduate School of Physical Sciences, Department of Physics,
Hiroshima University, Kagamiyama 1-3-1, Higashi-hiroshima, Hiroshima 739-8526, Japan}

\begin{abstract} 
The Minkowski vacuum state is expressed as an entangled state 
between the left and right Rindler wedges
when it is constructed on the Rindler vacuum. 
In this paper, we further examine the entanglement structure and extend the expression  
to the future (expanding) and past (shrinking) Kasner spacetimes.
This clarifies the origin of the quantum radiation produced by an Unruh--DeWitt detector 
in uniformly accelerated motion in the four-dimensional Minkowski spacetime.  
We also investigate the two-dimensional massless case where the quantum radiation vanishes
but the same entanglement structure exists.
\end{abstract} 

\maketitle

\def\I{{\rm I}}
\def\R{{\rm R}}
\def\II{{\rm II}}
\def\III{{\rm III}}
\def\IV{{\rm IV}}
\def\L{{\rm L}}
\def\F{{\rm F}}
\def\P{{\rm P}}
\def\d{{\rm d}}
\def\s{{\rm s}}

\begin{widetext}
\section{Introduction}
It is well known that the Minkowski vacuum state of a quantum field is described by an entangled state
between the left (L) and right (R) Rindler wedges when it is constructed on the Rindler vacuum \cite{UnruhWald}:
\begin{eqnarray}
\hspace{-5mm}
  |0,{\rm M}\rangle \propto \prod_j\Bigl[
\sum_{n_j=0}^\infty e^{-\pi n_j \omega_j /a } |n_j \rangle_{\rm R} \otimes|n_j\rangle_{\rm L}
\Bigr]  ,
\label{RLentangle}
\end{eqnarray}
where $|n_j \rangle_{\rm R(L)}$ is the $n_j$th excited state on the R (L) Rindler vacuum
with an acceleration $a$.
Because any physical quantity in the R Rindler wedge is not affected by  
the states in the left wedge, we can safely integrate them out and obtain a mixed state of
thermal equilibrium at the Unruh temperature $T_U=a/2\pi$  \cite{Unruh,UnruhWald}.
Consequently, the Unruh effect is usually considered to be a thermal phenomenon induced by 
quantum entanglement. 
The Unruh effect leads to various interesting theoretical predictions (see \cite{Higuchi} for a review) 
and plays a pivotal role in understanding phenomena in a system with a 
horizon, such as Hawking radiation in the black hole geometry or 
particle creation in the de Sitter universe. 

To demonstrate the Unruh effect, various experiments have been proposed 
\cite{Bell1,Bel2,SokolovTernov,SokolovTernov2,BarberMane,Unruh1998,Muller,Vanzella,SuzukiYamada,Vanzella2}.
One example is the quantum radiation emanating from a uniformly accelerated 
charged particle, which is called Unruh radiation 
\cite{ChenTajima,Schutzhold,Schutzhold2,ELI,Cozzella,LH,LH2,IYZ,Lin}.
The question of whether a uniformly accelerated object emits quantum radiation was studied
 for a two-dimensional case \cite{Grove,Raine}. 
It was soon confirmed that there is no radiation flux in a toy model 
of a uniformly accelerated detector in two-dimensional spacetime \cite{HuRaval}.
Massar et al. \cite{Massar} pointed out 
that a uniformly accelerated object generates a polarization cloud around the object,
but no radiation. 
The result is consistent with the intuition that the total flux of radiation is canceled
between outgoing and incoming fluxes from a thermal equilibrium system. 
However, the situation is different in the four-dimensional case. Lin and Hu reported that
a uniformly accelerated Unruh--DeWitt detector in four dimensions emits a 
positive radiated power of quantum radiation~\cite{LH,LH2}. 
Ref.~\cite{IOTYZ} showed that the total radiation flux is not canceled out,
and quantum radiation actually exists in a four-dimensional toy model, which 
confirmed the result of Ref.~\cite{LH,LH2}. 
This result was further generalized to a uniformly accelerated charged particle in 
four-dimensional spacetime \cite{IYZ,OYZ15,OYZ16}: Unruh radiation actually exists.

The presence of the radiation in 
the four-dimensional calculations seems to contradict the intuition that a system in thermal equilibrium
never emits radiation. In our previous papers \cite{IOTYZ, ITUY}, we 
pointed out that the apparent contradiction can be resolved by considering
the entanglement structure of the Minkowski vacuum in the future wedge.
Namely, in the future region to which most of the flux of quantum radiation propagates, 
the L Rindler states cannot be integrated out, and 
interference between the L and R Rindler wedge states in  (\ref{RLentangle}) 
 generates quantum Unruh radiation.
Thus, Unruh radiation is interpreted as entanglement-induced quantum radiation.

In this paper, we examine the entanglement structure of the Minkowski vacuum state
by extending the expression (\ref{RLentangle}) into the future (F) and past (P) degenerate Kasner universes 
shown in Fig. 1, expanding the work of Ref.~\cite{Higuchi}.
Equation (\ref{RLentangle}) can be extended by first rewriting the mode functions
in the Rindler wedges in terms of the wave functions defined globally 
in the entire Minkowski spacetime and then restricting them to the F and P Kasner universes. 
In Appendix A, we show that 
it can also be extended by using analytical continuations of the mode functions
from the L and R to the F and P regions across the horizons. 
By using the extension of the mode functions into the entire Minkowski spacetime, 
the expression (\ref{RLentangle}) is extended to 
\begin{eqnarray}
\hspace{-5mm}
  |0,{\rm M}\rangle \propto \prod_j\Bigl[
\sum_{n_j=0}^\infty e^{-\pi n_j \omega_j/a }|n_j,{\I}\rangle \otimes|n_j,{\II}\rangle
\Bigr]  .
\label{MIII0}
\end{eqnarray} 
Note that the wave functions representing the states $|n_j,{\I}\rangle$ and $|n_j,{\II}\rangle$ 
are defined in the entire spacetime, including the future and past regions (see also \cite{OlsonRalph}).
Thus, the formula makes it possible to calculate the correlations between the operators of 
different regions, e.g., the operator in the R Rindler region and that in the F Kasner region. 
It is even possible to calculate the correlations between the operators in the P and F regions. 
Deriving formula (\ref{MIII0}) is the main purpose of this paper.

Another purpose is to give the full details of the calculations in \cite{ITUY}.
We calculate the quantum radiation produced by a uniformly accelerated object.
In previous works~\cite{LH,LH2}, the radiation was derived using the Green function 
method with reference frame coordinates.
In contrast, to understand the physical origin of the radiation, 
we used a formalism based on the expression (\ref{MIII0}) of the Minkowski vacuum state.
It shows that quantum radiation is induced by entanglement of the vacuum
between the F region and the R Rindler region. 

This paper is organized as follows. In Sec.~II, we introduce the mode functions
of a massless scalar field in four-dimensional spacetime in each of the four 
regions, the R and L Rindler wedges and the F (expanding) and 
P (shrinking) degenerate Kasner spacetimes. 
In Sec.~III, we show how the expression (\ref{RLentangle}) for the Minkowski vacuum state
is extended to the F and P regions as in (\ref{MIII0}).
In Sec.~IV, a similar calculation is presented for two-dimensional spacetime 
for comparison to the four-dimensional case.
In Sec.~V, as an application of the entanglement structure studied in Sec.~III, 
we calculate the quantum radiation produced by a uniformly accelerated object.
The section explains the detailed calculations omitted in our 
previous letter \cite{ITUY}. 
In Sec.~VI, we study a similar system in two-dimensional spacetime \cite{HuRaval}.
In this case, the quantum radiation vanishes. This stems from a behavior of the mode functions
specific to the two-dimensional massless fields.
In Sec.~VII, a summary and conclusions are presented. 
In Appendix A, we give a different derivation of the results in Sec.~III
using analytical continuations of the mode functions across the horizons, expanding the work in 
Ref.~\cite{Sommerfield}. 
In Appendices B and C, calculations supplementing that in Sec.~V are given.

\section{Mode functions of four-dimensional massless fields in the {\rm R},~{\rm L},~{\rm F},~{\rm P} regions}
In this section, we first review the mode functions of a four-dimensional massless field in various
coordinate systems \cite{Higuchi}. 
We consider a massless scalar field whose action is given by 
\begin{eqnarray}
&&S={1\over 2}\int d^4x\sqrt{-g}g^{\mu\nu}\partial_\mu \phi\partial_\nu \phi
\label{fourdaction}
\end{eqnarray}
and quantize it in the following coordinate systems: the R Rindler 
wedge (R region), the L Rindler wedge (L region), the F (expanding) degenerate Kasner 
universe (F region), and the P (shrinking) degenerate Kasner universe (P region)
as well as the global Minkowski coordinates 
 (see Fig. \ref{fig:coordinate}).
The line element of the Minkowski spacetime is given by
\begin{eqnarray}
&&ds^2=dt^2-dz^2-d\bm x_\perp^2.
\label{fourdle}
\end{eqnarray}
Note that $\bm x_\perp$ denotes the two-dimensional coordinates perpendicular
to the $(t,z)$ plane.  
The equation of motion becomes
\begin{eqnarray}
&&\left({\partial^2\over \partial t^2}-{\partial^2\over \partial z^2}-
{\partial^2\over \partial \bm x_\perp^2}\right)\phi=0,
\label{fourdee}
\end{eqnarray}
and the quantized field is expanded as
\begin{eqnarray}
\phi&=&\int_{-\infty}^{\infty} {dk_zd^2k_\perp\over {(2\pi)}^{3/2}\sqrt{2k_0}}
\left( {\hat b}_{k_z\bm k_\perp}e^{-ik_0 t+ik_zz+i{\bm k}_\perp\cdot{\bm x}_\perp}
+{\rm h.c.}
\right)
, 
\end{eqnarray}
where the creation and annihilation operators satisfy the commutation relations
\begin{eqnarray}
[{\hat b}_{k_z\bm k_\perp}, {\hat b}_{k_z',\bm k_\perp'}^\dagger]=\delta_D(k_z-k_z') \delta_D^{(2)}(\bm k_\perp-\bm k_\perp'), 
~~
[{\hat b}_{k_z\bm k_\perp}, {\hat b}_{k_z',\bm k_\perp'}]=[{\hat b}_{k_z\bm k_\perp}^\dagger, {\hat b}_{k_z',\bm k_\perp'}^\dagger]=0. 
\end{eqnarray}
Here we defined $k_0=\sqrt{k_z^2+\bm k_\perp^2}$.
The Minkowski vacuum state $|0,{\rm M}\rangle$ is given by 
\begin{eqnarray}
&&\hat b_{k_z\bm k_\perp}|0,{\rm M}\rangle=0 ~~
\label{fourdvacuum}
\end{eqnarray}
for~any~$(k_z,~\bm k_\perp)$.

\begin{figure}[t]
\begin{center}
    \includegraphics[width=8cm]{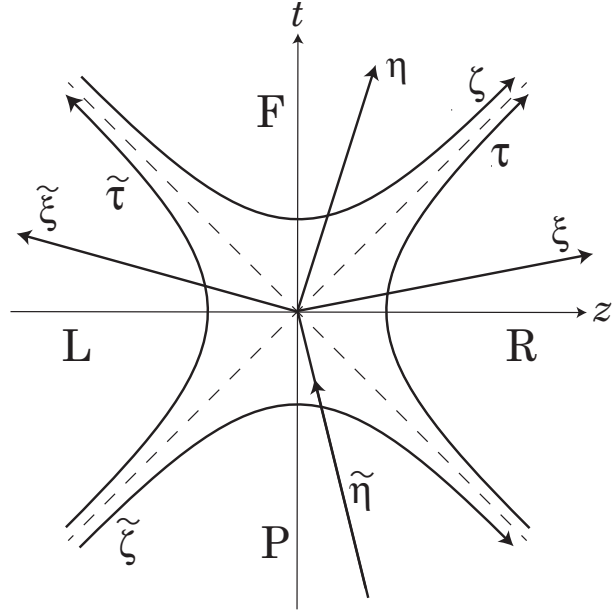}
\caption{Four regions of Minkowski spacetime and corresponding coordinates.
\label{fig:coordinate}}
\end{center}
\end{figure}

\subsection{R Rindler wedge (R region: $z>|t|$)}
The R Rindler wedge (R region) is described by the coordinates $\tau$ and $\xi$, 
\begin{eqnarray}
t={1\over a}e^{a\xi}\sinh a\tau,~~
z={1\over a}e^{a\xi}\cosh a\tau,
\end{eqnarray}
where $\tau$ and $\xi$ take values in the ranges $-\infty<\tau <\infty$ and $-\infty<\xi<\infty$.
These coordinates cover one-quarter of Minkowski spacetime (see Fig.~\ref{fig:coordinate}). 
The line element (\ref{fourdle}) and the field equation (\ref{fourdee}) reduce to
\begin{eqnarray}
ds^2=e^{2a\xi}(d\tau^2-d\xi^2)-d\bm x_\perp^2
\end{eqnarray}
and 
\begin{eqnarray}
&&\left({\partial^2\over \partial \tau^2}-{\partial^2\over \partial \xi^2}
-e^{2a\xi}{\partial^2\over \partial \bm x_\perp^2}\right)\phi=0,
\label{fourdeeR}
\end{eqnarray}
respectively. The quantized field can be expanded as
\begin{eqnarray}
\phi(x)=\int_{0}^{\infty} d \omega\int_{-\infty}^{\infty} d^2 k_\perp \left( {\hat a}_{\omega,{\bm k}_\perp}^{\rm R} v_{\omega,{\bm k}_\perp}^\R(x_\R)
+{\rm h.c.}
\right),
\label{qfR}
\end{eqnarray}
with the mode functions given by (see, e.g., \cite{Higuchi})
\begin{eqnarray}
v_{\omega,{\bm k}_\perp}^\R(x_\R)=\sqrt{\sinh \pi \omega/a\over 4\pi^4 a}
K_{i\omega/a}\left({\kappa e^{a\xi}\over a}\right)e^{i\bm k_\perp\cdot \bm x_\perp-i\omega\tau}.
\label{mfR}
\end{eqnarray}
The coordinate $\tau$ is the proper time of a uniformly accelerated observer at $\xi=0$,
and the mode function ($\omega>0$) represents a positive-frequency mode. 
Because the modified Bessel function 
$K_{i\omega/a}\left({\kappa e^{a\xi}\over a}\right)$ is approximated for 
 $\xi \rightarrow - \infty$ as 
\begin{eqnarray}
K_{i\omega/a}\left({\kappa e^{a\xi}\over a}\right) \approx
\frac{1}{2}\left( \frac{\kappa}{2a} \right)^{i \omega/a} \Gamma(\frac{-i\omega}{a}) \ e^{i \omega \xi} + {\rm c.c.},
\end{eqnarray}
the mode function $v_{\omega,{\bm k}_\perp}^\R(x_\R)$ represents 
a standing plane wave near the horizon, $\xi \rightarrow -\infty$. At $\xi \rightarrow +\infty$, 
it is rapidly damped because of the potential like $ \exp[-\kappa e^{2a\xi}/a]$, 
where we defined $\kappa=\sqrt{|\bm k_\perp|^2}$. 
The creation and annihilation operators satisfy the commutation relations
$
[{\hat a}^{\rm R}_{\omega,\bm k_\perp}, {\hat a}_{\omega',\bm k_\perp'}^{\rm R \dagger}]
=\delta_D(\omega-\omega') \delta_D^{(2)}(\bm k_\perp-\bm k_\perp'), 
~~
[{\hat a}^{\rm R}_{\omega,\bm k_\perp}, {\hat a}_{\omega',\bm k_\perp'}^{\rm R}]
=[{\hat a}^{\rm R \dagger}_{\omega,\bm k_\perp}, {\hat a}_{\omega',\bm k_\perp'}^{\rm R \dagger}]
=0.
$
The R Rindler vacuum state is defined as 
\begin{eqnarray}
&&
\hat a_{\omega,\bm k_\perp}^{\rm R} |0,{\rm R}\rangle=0
\label{fourdRvauum}
\end{eqnarray}
for any $(\omega, \bm k_\perp)$, and the $n_j$th excited R Rindler state is defined as 
\begin{eqnarray}
  &&|n_j,{\rm R}\rangle={1\over\sqrt{n_j!}}(\hat a_{j}^{\rm R \dagger})^{n_j}|0,{\rm R}\rangle,
\label{fourdRexcite}
\end{eqnarray}
where $j$ denotes the model specified by $j=(\omega,\bm k_\perp)$.\footnote{
Eq.~(\ref{fourdRexcite}) is a schematic expression, where some discretization of 
the modes is assumed with the normalization $[\hat a_j,\hat a_j^\dagger]=1$.}

\subsection{L Rindler wedge (L region: $-z>|t|$)}
The L Rindler wedge (L region in Fig.~\ref{fig:coordinate}) 
is similarly described by the coordinates $\tilde \tau$ and $\tilde\xi$, 
\begin{eqnarray}
t={1\over a}e^{a\tilde \xi}\sinh a\tilde \tau,~~
z=-{1\over a}e^{a\tilde \xi}\cosh a\tilde \tau,
\end{eqnarray}
where $\tilde \tau$ and $\tilde\xi$ take values in the ranges $-\infty<\tilde\tau<\infty$ and 
$-\infty<\tilde\xi<\infty$. The line element (\ref{fourdle}) becomes
\begin{eqnarray}
ds^2=e^{2a\tilde\xi}(d\tilde\tau^2-d\tilde\xi^2)-d\bm x_\perp^2.
\end{eqnarray}
The field equation is in the same form as (\ref{fourdee}), and 
we can expand the quantized field as
\begin{eqnarray}
\phi(x)=\int_{0}^{\infty} d \omega\int_{-\infty}^{\infty} d^2 k_\perp \left( {\hat a}_{\omega,{\bm k}_\perp}^{\rm L} v_{\omega,{\bm k}_\perp}^\L(x_\L)
+{\rm h.c.}
\right)
\label{qfL}
\end{eqnarray}
with the mode function
\begin{eqnarray}
v_{\omega,{\bm k}_\perp}^\L(x_\L)=\sqrt{\sinh \pi \omega/a\over 4\pi^4 a}
K_{i\omega/a}\left({\kappa e^{a\widetilde\xi}\over a}\right)e^{-i\bm k_\perp\cdot \bm x_\perp-i\omega\widetilde\tau},
\label{mfL}
\end{eqnarray}
where the creation and annihilation operators satisfy the commutation relations
$[{\hat a}^{\rm L}_{\omega,\bm k_\perp}, {\hat a}_{\omega',\bm k_\perp'}^{\rm L \dagger}]
=\delta_D(\omega-\omega') \delta_D^{(2)}(\bm k_\perp-\bm k_\perp'), 
~~
[{\hat a}^{\rm L}_{\omega,\bm k_\perp}, {\hat a}_{\omega',\bm k_\perp'}^{\rm L}]
=[{\hat a}^{\rm L \dagger}_{\omega,\bm k_\perp}, {\hat a}_{\omega',\bm k_\perp'}^{\rm L \dagger}]
=0
$. 
The L Rindler vacuum state is defined as 
$
\hat a_{\omega,\bm k_\perp}^{\rm L}|0,{\rm L}\rangle=0 
$
for~any $(\omega, \bm k_\perp)$, 
and the L Rindler particle state is defined as (see footnote 1) 
\begin{eqnarray}
&&|n_j,{\rm L}\rangle={1\over\sqrt{n_j!}}(\hat a_{j}^{\rm L \dagger})^{n_j}|0,{\rm L}\rangle.
\label{fourdLexcite}
\end{eqnarray}

\subsection{F (expanding) degenerate Kasner universe (F region: $t>|z|$)}
In the F (expanding) degenerate Kasner universe
(F region in Fig.~\ref{fig:coordinate}), 
we can introduce the coordinates $\eta$ and $\zeta$ as
\begin{eqnarray}
t={1\over a}e^{a\eta}\cosh a\zeta,~~
z={1\over a}e^{a\eta}\sinh a\zeta,
\end{eqnarray}
where $\eta$ and $\zeta$ take values in the ranges $-\infty<\eta<\infty$ and 
$-\infty<\zeta<\infty$. Thus, the line element (\ref{fourdle}) and
the field equation (\ref{fourdee}) become
\begin{eqnarray}
ds^2=e^{2a\eta}(d\eta^2-d\zeta^2)-d\bm x_\perp^2
\end{eqnarray}
and
\begin{eqnarray}
&&\left({\partial^2\over \partial \eta^2}-{\partial^2\over \partial \zeta^2}
-e^{2a\eta}{\partial^2\over \partial \bm x_\perp^2}\right)\phi=0,
\label{fourdeF}
\end{eqnarray}
respectively. Because $\eta$ is the time variable, the metric describes an expanding universe with the 
scale factor $e^{a \eta}$.
 The quantized field is expanded as 
\begin{eqnarray}
\phi(x)=\int_{-\infty}^{\infty} 
d \omega\int_{-\infty}^{\infty} d^2k_\perp \left( {\hat a}_{\omega,{\bm k}_\perp}^\F v_{\omega,{\bm k}_\perp}^\F(x_\F)
+{\rm h.c.}
\right)
\label{qfF}
\end{eqnarray}
in terms of the mode functions 
\begin{eqnarray}
v_{\omega,{\bm k}_\perp}^\F(x_\F)={-ie^{i\omega\zeta}\over 2\pi\sqrt{4a\sinh (\pi |\omega|/a)}}
J_{-i|\omega|/a}\left({\kappa e^{a\eta}\over a}\right)
e^{i\bm k_\perp\cdot \bm x_\perp}.
\label{qfRR}
\end{eqnarray}
Because the Bessel function $J_{-i|\omega|/a}\left({\kappa e^{a\eta}\over a}\right)$ 
is approximated at $\eta \rightarrow -\infty$ (near the horizon) as
\begin{eqnarray}
J_{-i|\omega|/a}\left({\kappa e^{a\eta}\over a}\right) 
\propto  e^{-i |\omega| \eta},
\end{eqnarray}
it represents a positive-frequency mode for both positive and negative $\omega$.
Mode functions with a positive $\omega$  
represent right-moving wave modes in the $\zeta$ direction,
whereas modes with a negative
$\omega$ represent left-moving wave modes. 
The creation and annihilation operators satisfy the commutation relations
$[{\hat a}^\F_{\omega,\bm k_\perp}, {\hat a}_{\omega',\bm k_\perp'}^{F\dagger}]
=\delta_D(\omega-\omega') \delta_D^{(2)}(\bm k_\perp-\bm k_\perp'), 
~~
[{\hat a}^\F_{\omega,\bm k_\perp}, {\hat a}_{\omega',\bm k_\perp'}^\F]
=[{\hat a}^{F\dagger}_{\omega,\bm k_\perp}, {\hat a}_{\omega',\bm k_\perp'}^{F\dagger}]
=0.
$
We define the vacuum state in the F region as 
\begin{eqnarray}
&&
\hat a_{\omega,\bm k_\perp}^\F|0,{\rm F}\rangle=0 
\label{fourdFvauum}
\end{eqnarray}
for any $(\omega,~\bm k_\perp)$, and the excited particle states are also defined using 
$\hat a_{\omega,\bm k_\perp}^{F\dagger}$. 

Because a positive (negative) $\omega$ represents a right (left)-moving wave mode in the $\zeta$ 
direction, 
 we can decompose the field $\phi$  into 
\begin{eqnarray}
  &&\phi(x)=\phi^{\F,\d}(x)+\phi^{\F,\s}(x).
\label{phiFF}
\end{eqnarray}
Here we define the {\it sinister} (``left'' in Latin) field
\begin{eqnarray}
 &&\phi^{\F,\s}(x)= \int_{0}^{\infty} 
 d \omega\int_{-\infty}^{\infty} d^2 k_\perp \Bigl({\hat a}_{\omega,{\bm k}_\perp}^{\F,\s}v_{\omega,{\bm k}_\perp}^{\F,\s}(x)
+ {\rm h.c.}
\Bigr)
\end{eqnarray}
with the mode function 
\begin{eqnarray}
v_{\omega,{\bm k}_\perp}^{\F,\s}(x) :=v_{-\omega,{\bm k}_\perp}^\F(x)
=
{-ie^{-i\omega\zeta}\over 2\pi\sqrt{4a\sinh (\pi \omega/a)}}
J_{-i\omega/a}\left({\kappa e^{a\eta}\over a}\right)
e^{i\bm k_\perp\cdot \bm x_\perp},
\end{eqnarray}
which contains only the left-moving modes in the $\zeta$ direction near the horizon. 
In contrast, the {\it dexter} (``right'' in Latin) field, 
\begin{eqnarray}
\phi^{\F,\d}(x) = \int_{0}^{\infty} 
d \omega\int_{-\infty}^{\infty} d^2 k_\perp \Bigl( {\hat a}_{\omega,{\bm k}_\perp}^{\F,\d}v_{\omega,{\bm k}_\perp}^{\rm F,d}(x)
+{\rm h.c.}
\Bigr),
\end{eqnarray}
is defined with the mode function 
\begin{eqnarray}
v_{\omega,{\bm k}_\perp}^{\F,\d}(x):=v_{\omega,-{\bm k}_\perp}^\F(x)
=
{-ie^{i\omega\zeta}\over 2\pi\sqrt{4a\sinh (\pi \omega/a)}}
J_{-i\omega/a}\left({\kappa e^{a\eta}\over a}\right)
e^{-i\bm k_\perp\cdot \bm x_\perp},
\end{eqnarray}
which
contains only the right-moving modes in the $\zeta$ direction near the horizon. 
The annihilation operators are defined accordingly as 
${\hat a}_{\omega,{\bm k}_\perp}^{\F,\s}={\hat a}_{-\omega,{\bm k}_\perp}^{\F}$
and 
${\hat a}_{\omega,{\bm k}_\perp}^{\F,\d}={\hat a}_{\omega,-{\bm k}_\perp}^{\F}$.
In both expansions of the field, $\omega$ takes a positive value. 

\subsection{P (shrinking) degenerate Kasner universe (P region: $-t>|z|$)}
In the P (shrinking) degenerate Kasner universe (P region in Fig.~\ref{fig:coordinate}),
we introduce the coordinates $\tilde\eta$ and $\tilde\zeta$ 
\begin{eqnarray}
t=-{1\over a}e^{-a\tilde\eta}\cosh a\tilde\zeta,~~
z={1\over a}e^{-a\tilde\eta}\sinh a\tilde\zeta.
\end{eqnarray}
These variables, $\tilde\eta$ and $\tilde\zeta$, take values in the ranges $-\infty<\tilde\eta<\infty$ and 
$-\infty<\tilde\zeta<\infty$. The line element (\ref{fourdle}) and
the field equation (\ref{fourdee}) become
\begin{eqnarray}
ds^2=e^{-2a\tilde\eta}(d\tilde\eta^2-d\tilde \zeta^2)-d\bm x_\perp^2
\end{eqnarray}
and 
\begin{eqnarray}
&&\left({\partial^2\over \partial \tilde\eta^2}-{\partial^2\over \partial \tilde\zeta^2}
-e^{-2a\tilde\eta}{\partial^2\over \partial \bm x_\perp^2}\right)\phi=0,
\label{fourdeP}
\end{eqnarray}
respectively. The quantized field is expanded as
\begin{eqnarray}
  \phi(x) &=&\int_{-\infty}^{\infty} d \omega\int_{-\infty}^{\infty} d^2 k_\perp \left( {\hat a}_{\omega,{\bm k}_\perp}^\P
  v_{\omega,{\bm k}_\perp}^\P(x_\P)
+{\rm h.c.}
\right),
\label{qfP}
\end{eqnarray}
where the mode function is defined as
\begin{eqnarray}
v_{\omega,{\bm k}_\perp}^\P(x_\P)={ie^{i\omega \tilde\zeta}\over 2\pi\sqrt{4a\sinh (\pi |\omega|/a)}}
J_{i|\omega|/a}\left({\kappa e^{-a\tilde\eta}\over a}\right)
e^{i\bm k_\perp\cdot \bm x_\perp}.
\label{qfPP}
\end{eqnarray}
The creation and annihilation operators satisfy the commutation relations
$[{\hat a}^\P_{\omega,\bm k_\perp}, {\hat a}_{\omega',\bm k_\perp'}^{\P\dagger}]
=\delta_D(\omega-\omega') \delta_D^{(2)}(\bm k_\perp-\bm k_\perp'), 
~~
[{\hat a}^\P_{\omega,\bm k_\perp}, {\hat a}_{\omega',\bm k_\perp'}^\P]
=[{\hat a}^{\P\dagger}_{\omega,\bm k_\perp}, {\hat a}_{\omega',\bm k_\perp'}^{\P\dagger}]
=0$. 
The vacuum state $|0,{\rm P}\rangle$ in the P region is defined as 
\begin{eqnarray}
&&
\hat a_{\omega,\bm k_\perp}^\P|0,{\rm P}\rangle=0
\label{fourdPvauum}
\end{eqnarray}
for any $(\omega,~\bm k_\perp)$, and the excited particle states are created 
using the operators $\hat a_{\omega,\bm k_\perp}^{\P\dagger}$. 

The $\omega$ in (\ref{qfPP}) is the momentum 
in the $\tilde\zeta$ direction, and a positive (negative) $\omega$ represents a right (left)-moving wave mode. 
Thus, we can separate the right-moving wave modes from the left-moving wave modes by decomposing
\begin{eqnarray}
  &&\phi(x)=\phi^{\P,\d}(x)+\phi^{\P,\s}(x),
\label{phiPP}
\end{eqnarray}
where we define
\begin{eqnarray}
 &&\phi^{\P,\s}(x)= \int_{0}^{\infty} 
 d \omega\int_{-\infty}^{\infty} d^2 k_\perp \Bigl({\hat a}_{\omega,{\bm k}_\perp}^{\P,\s}v_{\omega,{\bm k}_\perp}^{\P,\s}(x)
+ {\rm h.c.}
\Bigr),
\nonumber
\\
&&\phi^{\P,\d}(x)= \int_{0}^{\infty} 
d \omega\int_{-\infty}^{\infty} d^2 k_\perp \Bigl( {\hat a}_{\omega,{\bm k}_\perp}^{\P,\d}
v_{\omega,{\bm k}_\perp}^{\P,\d}(x)
+{\rm h.c.}
\Bigr).
\nonumber
\end{eqnarray}
The mode functions are defined as 
\begin{eqnarray}
&&\hspace{-5mm}
v_{\omega,{\bm k}_\perp}^{\P,\s}(x)=v_{-\omega,-{\bm k}_\perp}^\P(x)
={ie^{-i\omega \tilde\zeta}\over 2\pi\sqrt{4a\sinh (\pi \omega/a)}}
J_{i\omega/a}\left({\kappa e^{-a\tilde\eta}\over a}\right)
e^{-i\bm k_\perp\cdot \bm x_\perp},
\label{vps}
\\
  &&\hspace{-5mm}
  v_{\omega,{\bm k}_\perp}^{\P,\d}(x)=v_{\omega,{\bm k}_\perp}^\P(x)
 ={ie^{i\omega \tilde\zeta}\over 2\pi\sqrt{4a\sinh (\pi \omega/a)}}
J_{i\omega/a}\left({\kappa e^{-a\tilde\eta}\over a}\right)
e^{i\bm k_\perp\cdot \bm x_\perp},
\label{vpd}
\end{eqnarray}
and the annihilation operators are defined accordingly as
\begin{eqnarray}
&&{\hat a}_{\omega,{\bm k}_\perp}^{\P,\s}={\hat a}_{-\omega,-{\bm k}_\perp}^{\P},
\\
&&{\hat a}_{\omega,{\bm k}_\perp}^{\P,\d}={\hat a}_{\omega,{\bm k}_\perp}^{\P}.
\end{eqnarray}

\section{Description of the Minkowski vacuum state }
In this section, we will connect the mode functions defined in each coordinate system
in the previous section and express the Minkowski vacuum state 
as in (\ref{MIII0}), which is defined in the entire Minkowski spacetime.
\subsection{Positive-frequency modes in Minkowski spacetime  \label{3A}}
First, we note that the following 
linear combinations of the mode functions in the R and L regions (normalized with respect to the Klein--Gordon inner product) are positive-frequency modes in the Minkowski sense (see, e.g., Ref.~\cite{Higuchi}):\footnote{Here,
$v_{\omega{\bm k}_\perp}^{\L}$ in Ref.~\cite{Higuchi} is $v_{\omega,-{\bm k}_\perp}^{\L}$ in the present paper.}
\begin{eqnarray}
w_{- \omega,{\bm k}_\perp}
& = & \frac{v^\R_{\omega, {\bm k}_\perp} + e^{-\pi\omega/a}v_{\omega,{\bm k}_\perp}^{\L*}}{\sqrt{ 1 - e^{-2\pi\omega/a}}},
\label{eqfivea}
\\
w_{+\omega,{\bm k}_\perp}
& = & \frac{v^\L_{\omega,-{\bm k}_\perp} + e^{-\pi\omega/a}v_{\omega,-{\bm k}_\perp}^{\R*}}{\sqrt{1-e^{-2\pi \omega/a}}} .
\label{eqsixa}
\end{eqnarray}
As shown in \cite{Higuchi}, 
these positive-frequency modes in Minkowski spacetime are written in the integral form
\begin{eqnarray}
w_{\pm\omega,\bm k_\perp}=\int_{-\infty}^{\infty} {dk_z\over \sqrt{8a}\pi k_0}e^{\pm i\theta(k_z)\omega/a}
e^{-ik_0t+ik_zz}{e^{i\bm k_\perp\cdot \bm x_\perp}\over 2\pi},
\label{definitionw}
\end{eqnarray}
where the rapidity variable $\theta$ is defined as $k_0=\kappa \cosh\theta$, $k_z=\kappa \sinh\theta$.
It is thus written as 
\begin{eqnarray}
\theta(k_z)={1\over 2}\ln\left({k_0+k_z\over k_0-k_z}\right).
\label{definitiontheta}
\end{eqnarray}
We introduce the (unnormalized) positive-frequency modes in Minkowski spacetime $W_{\pm \omega}$ as
\begin{eqnarray}
W_{\pm \omega}=\int_{-\infty}^{\infty} {d\theta }e^{\pm i\theta\omega/a} e^{-i\kappa (t\cosh \theta-z\sinh\theta)}
\label{definitionWW}
\end{eqnarray}
and evaluate $W_{\pm \omega}$ in the L (R) Rindler and F(P) Kasner coordinates. 
Note that $W_{\pm \omega}$ is related to $w_{\pm\omega,\bm k_\perp}$ by
 \begin{eqnarray}
w_{\pm\omega,\bm k_\perp}={e^{i\bm k_\perp\cdot \bm x_\perp}\over \sqrt{2a}(2\pi)^2} W_{\pm \omega}.
\end{eqnarray}
For $\bm k_\perp=0$, we cannot write the 
mode functions in the above form [(\ref{definitionWW})] because of the relation $k_0=|k_z|$. 
This case is investigated separately in Sec.~IV as a two-dimensional massless field. 
In the four-dimensional case, the modes with $\bm k_\perp=0$ 
make negligible contributions to the composition of a physically relevant wave 
packet.

Eqs. (\ref{eqfivea}) and (\ref{eqsixa}) are solved to obtain
the R Rindler mode $v_{\omega,\bm k_\perp}^{\R}$ and
L Rindler mode $v_{\omega,\bm k_\perp}^{\L}$ in terms of the positive-frequency Minkowski
modes. Renaming them as $v_{\omega,\bm k_\perp}^{\I}$ and $v_{\omega,\bm k_\perp}^{\II}$ for later
convenience,
we have
\begin{eqnarray}
v_{\omega,\bm k_\perp}^{\R} &\rightarrow& v^{\rm I}_{\omega,{\bm k}_\perp}
=  \frac{w_{-\omega, {\bm k}_\perp} - e^{-\pi\omega/a}w_{+\omega,-{\bm k}_\perp}^{*}}{\sqrt{ 1 - e^{-2\pi\omega/a}}},
\label{eqfiveai}
\\
v_{\omega,\bm k_\perp}^{\L} &\rightarrow&v^{\rm II}_{\omega,{\bm k}_\perp}
 =  \frac{w_{+\omega,-{\bm k}_\perp} - e^{-\pi\omega/a}w_{-\omega,{\bm k}_\perp}^{*}}{\sqrt{1-e^{-2\pi \omega/a}}} .
\label{eqsixai}
\end{eqnarray}
These wave functions are 
defined in the entire Minkowski spacetime through the integral representation of (\ref{definitionWW}).
This makes a striking contrast to the original definitions of $v_{\omega,\bm k_\perp}^{\R}$
and $v_{\omega,\bm k_\perp}^{\L}$, which are defined only in the restricted regions, namely,
in the R and L Rindler wedges, respectively.

After briefly discussing the R and L regions for clarity, 
we explicitly evaluate the integral in the F and P regions and relate
  $v_{\omega,\bm k_\perp}^{\I}$ and $v_{\omega,\bm k_\perp}^{\II}$
 with the mode functions defined there.

\subsection{R region}
In the R region, (\ref{definitionWW}) reduces to 
\begin{eqnarray}
W_{\pm \omega}&=&\int_{-\infty}^{\infty} {d\theta }e^{\pm i\theta\omega/a} e^{i\kappa (e^{a\xi}/a)\sinh (\theta-a\tau) }
=e^{\pm i\omega\tau}\int_{-\infty}^{\infty} {d\theta }e^{\pm i\theta\omega/a} e^{i\kappa (e^{a\xi}/a)\sinh \theta}.
\label{definitionWR}
\end{eqnarray}
Considering the convergence property at $|\theta|=\infty$ on the complex plane, 
the integration contour of $\theta$ can be shifted to
the contour $\theta= x +\pi i/2-i\varepsilon $ with $(\varepsilon>0)$ 
in the range $-\infty < x < \infty$:
\begin{eqnarray}
W_{\pm \omega}&=&e^{\pm i\omega\tau\mp \pi\omega/2a}\int_{-\infty}^{\infty} {dx }e^{\pm ix\omega/a} 
e^{-(\kappa e^{a\xi}/a)\cosh x} ,
\label{definitionWR2}
\end{eqnarray}
where we used $\sinh(x+\pi i/2)=i\cosh x$. 
Because the integral representation of the modified Bessel function is given by
\begin{eqnarray}
K_{i\omega}(z)=\int_{-\infty}^{\infty} {dx }e^{\pm x\nu-z\cosh t}  
e^{-(\kappa e^{a\xi}/a)\cosh x},
\label{definitionK}
\end{eqnarray}
we have 
\begin{eqnarray}
&&v^{\I}_{\omega,\bm k_\perp}=e^{-i\omega\tau} e^{i\bm k_\perp\cdot \bm x_\perp}\sqrt{\sinh \pi\omega/a \over 4\pi^4 a}
K_{i\omega/a}\left({\kappa e^{a\xi} \over a}\right)=v^{\R}_{\omega,\bm k_\perp}(x_{\rm R}),
\label{I=R}
\end{eqnarray}
and
\begin{eqnarray}
&&v^{\II}_{\omega,\bm k_\perp}=0 .
\label{II=0inR}
\end{eqnarray}
The result in (\ref{I=R}) just confirms (\ref{eqfiveai}).
On the other hand, (\ref{II=0inR}) shows that the wave function defined in the L 
Rindler wedge (\ref{eqsixai}) has no support in the R Rindler wedge.

\subsection{L region}
In the L region, (\ref{definitionWW}) reduces to 
\begin{eqnarray}
W_{\pm \omega}&=&\int_{-\infty}^{\infty} {d\theta }e^{\pm i\theta\omega/a} e^{-i\kappa (e^{a\tilde\xi}/a)\sinh (\theta+a\tilde\tau) }.
\label{definitionWL}
\end{eqnarray}
By changing the contour on the complex plane, as in the R region,
it is written as
\begin{eqnarray}
W_{\pm \omega}&=&e^{\mp i\omega\tilde\tau\pm \pi\omega/2a}\int_{-\infty}^{\infty} {dx }e^{\mp ix\omega/a} 
e^{-(\kappa e^{a\tilde\xi}/a)\cosh x},
\label{definitionWR2}
\end{eqnarray}
and $v^{\rm I, II}_{\omega, {\bf k}_\perp}$ are evaluated in the L region as
\begin{eqnarray}
&&v^{\I}_{\omega,\bm k_\perp}=0,
\label{I=0inL}
\\
&&v^{\II}_{\omega,\bm k_\perp}=e^{-i\omega\tilde\tau} e^{-i\bm k_\perp\cdot \bm x_\perp}\sqrt{\sinh \pi\omega/a \over 4\pi^4 a}
K_{i\omega/a}\left({\kappa e^{a\tilde \xi} \over a}\right)
=v^{\L}_{\omega,\bm k_\perp}(x_{\rm L}).
\label{II=L}
\end{eqnarray}
The result in (\ref{II=L}) just confirms (\ref{eqsixai}).
On the other hand, (\ref{I=0inL}) shows that the wave function defined in the R 
Rindler wedge (\ref{eqfiveai}) has no support in the L Rindler wedge.
\begin{figure}[b]
\begin{center}
    \includegraphics[width=6.cm]{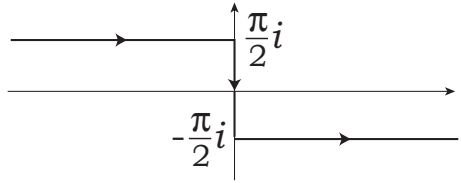}
\caption{Contour of integration on the complex plane.
\label{CPcont}}
\end{center}
\end{figure}
\subsection{F region}
In the F region, 
 (\ref{definitionWW}) reduces to 
\begin{eqnarray}
W_{\pm \omega}&=&\int_{-\infty}^{\infty} {d\theta }e^{\pm i\theta\omega/a} e^{-i\kappa (e^{a\eta}/a)\cosh (\theta-a\zeta) }
=e^{\pm i\omega\zeta}\int_{-\infty}^{\infty} {d\theta }e^{\pm i\theta\omega/a} e^{-i\kappa (e^{a\eta}/a)\cosh \theta }.
\label{definitionWF}
\end{eqnarray}
In this case, because the convergence property at $|\theta|=\infty$ is different from that in the R Rindler case,
 the integration contour can be shifted on the complex plane, as shown in Fig.~\ref{CPcont}. 
Then the integration is rewritten as
\begin{eqnarray}
W_{\pm \omega}&=&e^{\pm i\omega\zeta}\biggl\{
e^{\mp \pi\omega/2a}\int_0^\infty dx e^{\mp ix\omega/a} e^{-(\kappa e^{a\eta}/a)\sinh x}
+e^{\pm \pi\omega/2a}\int_0^\infty dx e^{\pm ix\omega/a} e^{-(\kappa e^{a\eta}/a)\sinh x}
\nonumber
\\
&&+\int_{\pi i/2}^{-\pi i/2} d\theta e^{\pm i\theta \omega/a} e^{-i(\kappa e^{a\eta}/a)\cosh\theta}\biggr\}.
\end{eqnarray}
The last term on the right-hand side of the above equation reduces to 
\begin{eqnarray}
-i e^{\pm i\omega\zeta} e^{\pm \pi \omega/2a}\int _0^\pi dx e^{\mp \omega x/a} e^{-iz\sin x}. 
\end{eqnarray}
By using the Bessel Schl\"afli integration formula 
\begin{eqnarray}
J_\nu(z)={1\over \pi} \int_0^\pi \cos(\nu t-z\sin t) dt-{\sin\nu\pi\over \pi}
\int_0^\infty dt e^{-\nu t-z\sinh t},
\end{eqnarray}
we find that 
\begin{eqnarray}
&&v^{\I}_{\omega,\bm k_\perp}=
{-ie^{-i\omega\zeta}\over 2\pi\sqrt{4a\sinh (\pi \omega/a)}}
J_{-i\omega/a}\left({\kappa e^{a\eta}\over a}\right)
e^{i\bm k_\perp\cdot \bm x_\perp}=v_{\omega,{\bm k}_\perp}^{\F,\s}(x),
\\
&&v^{\II}_{\omega,\bm k_\perp}=
{-ie^{i\omega\zeta}\over 2\pi\sqrt{4a\sinh (\pi \omega/a)}}
J_{-i\omega/a}\left({\kappa e^{a\eta}\over a}\right)
e^{-i\bm k_\perp\cdot \bm x_\perp}
=v_{\omega,{\bm k}_\perp}^{\F,\d}(x).
\end{eqnarray}
As shown in Fig.~\ref{fig:coordinate4},
the wave function $v^{\rm I}_{\omega,\bm k_\perp}$ is originally defined in the R region and vanishes in the L region.
Consequently, in the F region, it becomes the left-moving mode $v^{\rm F, s}_{\omega,\bm k_\perp}$
near the horizon between the R Rindler and F Kasner wedges.
On the other hand, the wave 
function $v^{\rm II}_{\omega,\bm k_\perp}$ is originally defined in the L region and vanishes in the R region.
Consequently, in the F region, it becomes the right-moving mode $v^{\rm F, d}_{\omega,\bm k_\perp}$
near the horizon between the L Rindler and F Kasner wedges.

\subsection{P region}
In the P region, 
(\ref{definitionWW}) reduces to 
\begin{eqnarray}
W_{\pm \omega}&=&\int_{-\infty}^{\infty} {d\theta }e^{\pm i\theta\omega/a} e^{i\kappa (e^{-a\tilde\eta}/a)\cosh (\theta+a\tilde\zeta) }
=e^{\mp i\omega\tilde\zeta}\int_{-\infty}^{\infty} {d\theta }e^{\pm i\theta\omega/a} e^{i\kappa (e^{-a\tilde\eta}/a)\cosh \theta }.
\label{definitionWP}
\end{eqnarray}
This is obtained by taking the complex conjugate of $W_{\pm \omega}$ in the F region and replacing 
$\eta$ and $\zeta$ with $-\tilde\eta$ and $\tilde\zeta$, respectively, and we find
\begin{eqnarray}
&&v^{\I}_{\omega,\bm k_\perp}=
{ie^{i\omega\tilde\zeta}\over 2\pi\sqrt{4a\sinh (\pi \omega/a)}}
J_{i\omega/a}\left({\kappa e^{-a\tilde\eta}\over a}\right)
e^{i\bm k_\perp\cdot \bm x_\perp}=v_{\omega,{\bm k}_\perp}^{\P,\d}(x),
\\
&&v^{\II}_{\omega,\bm k_\perp}=
{ie^{-i\omega\tilde\zeta}\over 2\pi\sqrt{4a\sinh (\pi \omega/a)}}
J_{i\omega/a}\left({\kappa e^{-a\tilde\eta}\over a}\right)
e^{-i\bm k_\perp\cdot \bm x_\perp}
=v_{\omega,{\bm k}_\perp}^{\P,\s}(x).
\end{eqnarray}
As shown in Fig.~\ref{fig:coordinate4},
the wave functions $v^{\rm I}$ ($v^{\rm II}$) 
appear in the P region as the right (left)-moving mode $v^{\rm P, d}$ ($v^{\rm P, s}$) 
near the horizon between the R (L) Rindler and P Kasner wedges.

\begin{figure}[t]
\begin{center}
    \includegraphics[width=7.5cm]{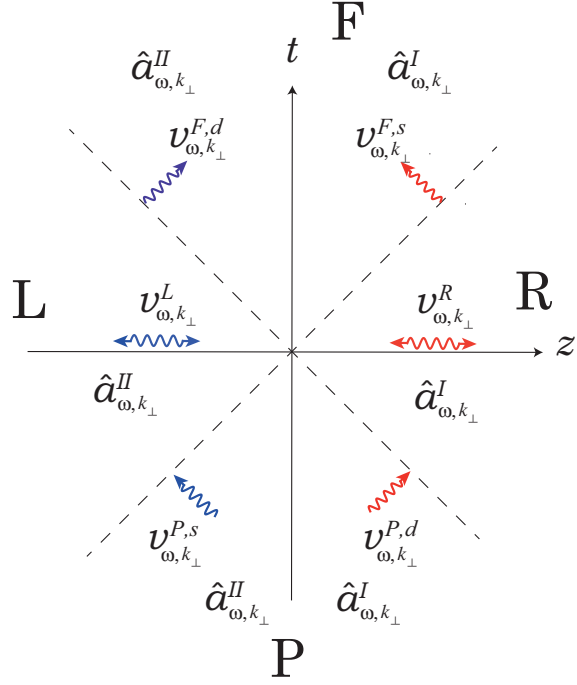}
\caption{Mode functions in each region and their relationships. 
The red and blue undulating modes are 
$v^\I_{\omega,\bm k_\perp}$ and $v^\II_{\omega,\bm k_\perp}$, respectively, 
which are associated with the annihilation (creation) operators 
$\hat a^\I_{\omega,\bm k_\perp}$ ($\hat a^{\I\dagger}_{\omega,\bm k_\perp}$) 
and $\hat a^{\II}_{\omega,\bm k_\perp}$ ($\hat a^{\II\dagger}_{\omega,\bm k_\perp}$). 
\label{fig:coordinate4}
}
\end{center}
\end{figure}

\subsection{Entanglement of the Minkowski vacuum state \label{Sec:entanglement}}
Summarizing the results in the previous subsections, we find that 
the functions $v^\I_{\omega,\bm k_\perp}$ and $v^\II_{\omega,\bm k_\perp}$
are identified with the functions introduced in Sec.~II as follows:
\begin{eqnarray}
v_{\omega,\bm k_\perp}^{\rm I}(x)=\left\{
\begin{array}{lc}
v_{\omega,{\bm k}_\perp}^{\F,\s} & ~~{\rm F}
\\
v^\R_{\omega,{\bm k}_\perp}& ~~{\rm R}
\\
0& ~~{\rm L}
\\
v_{\omega,{\bm k}_\perp}^{\P,\d} & ~~{\rm P}
\end{array}
\right.,
~~~~~~~~~~~~~
v_{\omega,\bm k_\perp}^{\rm II}(x)=\left\{
\begin{array}{lc}
v_{\omega,{\bm k}_\perp}^{\F,\d} & ~~{\rm F}
\\
0& ~~{\rm R}
\\
v^\L_{\omega,{\bm k}_\perp}& ~~{\rm L}
\\
v_{\omega,{\bm k}_\perp}^{\P,\s} & ~~{\rm P}
\end{array}
\right..
\end{eqnarray}
The behavior of the wave functions is drawn schematically in Fig.~\ref{fig:coordinate4}.
It can also be obtained by using continuations of the 
wave functions through the Minkowski positive-frequency mode functions (see Appendix A).  
Thus, we can obtain globally defined annihilation operators, $\hat a^{\rm I}_{\omega,\bm k_\perp}$
and $\hat a^{\rm II}_{\omega,\bm k_\perp}$, by 
making the following identifications:
\begin{eqnarray}
 &&\hat a^{\rm I}_{\omega,\bm k_\perp} := \hat a^{\rm R}_{\omega,\bm k_\perp}
=\hat a^{\P,\d}_{\omega,\bm k_\perp}
=\hat a^{\F,\s}_{\omega,\bm k_\perp},
\nonumber\\
 &&\hat a^{\rm II}_{\omega,\bm k_\perp} := \hat a^{\rm L}_{\omega,\bm k_\perp}
=\hat a^{\P,\s}_{\omega,\bm k_\perp}
=\hat a^{\F,\d}_{\omega,\bm k_\perp}.
\nonumber
\end{eqnarray}
We can now expand the quantum field in terms of the mode functions 
$v^\sigma_{\omega,\bm k_\perp}(x)$ ($\sigma = {\rm I, II})$ and the globally defined operators as
\begin{eqnarray}
&&\phi(x)=\sum_{\sigma={\rm I,II}}\int_0^\infty d\omega\int_{-\infty}^\infty d^2 k_\perp 
\left(
\hat a^\sigma_{\omega,\bm k_\perp}v^\sigma_{\omega,\bm k_\perp}(x)+{\rm h.c.}
\right) .
\label{mode-I-II}
\end{eqnarray}
Note that the expansion is valid in the entire Minkowski region. 
By using these modes, 
the Minkowski vacuum state is written as 
\begin{eqnarray}
\hspace{-5mm}
  |0,{\rm M}\rangle=\prod_j\Bigl[
N_j\sum_{n_j=0}^\infty e^{-\pi n_j \omega_j/a }|n_j,{\I}\rangle \otimes|n_j,{\II}\rangle
\Bigr]  ,
\label{MIII}
\end{eqnarray}
where $N_j=\sqrt{1-e^{-2\pi \omega_j /a}}$ and  $j=(\omega_j,\bm k_\perp)$. 
Notice that $\omega_j$ takes only positive values.

Eq.~(\ref{MIII}) expresses the Minkowski vacuum state in four-dimensional spacetime
as an entangled state constructed on the basis of the modes $v^{\rm I}$ and $v^{\rm II}$. 
The entanglement structure of the Minkowski vacuum between the R 
and L regions is well-known \cite{Unruh,UnruhWald}, and 
(\ref{MIII}) gives a generalization to the F and the P regions. 
Therefore, it is now possible to calculate the correlations between the operators of any of the regions 
R, L, F, and P.
For example, as discussed in Ref.~\cite{ITUY}, the entanglement between the states in the F
and R regions is important for understanding the quantum radiation associated with the Unruh effect.
This will be discussed in Sec.~\ref{Sec:Application1}.
It also represents timelike entanglement between states
in the F and P regions \cite{OlsonRalph}.

\subsection{Two-point Wightman function \label{Sec:2pointWightman} }
To confirm completeness of the mode expansion in (\ref{mode-I-II}), 
we compute the two-point Wightman function using (\ref{mode-I-II}) and show that it reproduces
the two-point function calculated in the Minkowski basis as in Ref.~\cite{Higuchi}.
The two-point Wightman function is computed as
\begin{eqnarray}
&&\langle0,{\rm M}|\phi(x)\phi(x')|0,{\rm M}\rangle=\int_0^\infty d\omega\int_{-\infty}^{\infty} 
d^2 k_\perp \biggl[\left\{
v_{\omega,{\bm k}_\perp}^{\rm I}(x)v_{\omega,{\bm k}_\perp}^{\rm I*}(x')
+v_{\omega,{\bm k}_\perp}^{\rm II}(x)v_{\omega,{\bm k}_\perp}^{\rm II*}(x')
\right\}{1\over 1-e^{-2\pi\omega/a}}
\nonumber\\
&&+\left\{
v_{\omega,{\bm k}_\perp}^{\rm I*}(x)v_{\omega,{\bm k}_\perp}^{\rm I}(x')
+v_{\omega,{\bm k}_\perp}^{\rm II*}(x)v_{\omega,{\bm k}_\perp}^{\rm II}(x')
\right\}{1\over e^{2\pi\omega/a}-1}
+\bigl\{
v_{\omega,{\bm k}_\perp}^{\rm I}(x)v_{\omega,{\bm k}_\perp}^{\rm II}(x')
+v_{\omega,{\bm k}_\perp}^{\rm I*}(x)v_{\omega,{\bm k}_\perp}^{\rm II*}(x')
\nonumber\\
&&+
v_{\omega,{\bm k}_\perp}^{\rm II}(x)v_{\omega,{\bm k}_\perp}^{\rm I}(x')
+v_{\omega,{\bm k}_\perp}^{\rm II*}(x)v_{\omega,{\bm k}_\perp}^{\rm I*}(x')
\bigr\}{e^{\pi\omega/a}\over e^{2\pi\omega/a}-1}\biggr],
\end{eqnarray}
where we used
\begin{eqnarray}
&& \langle0,{\rm M}|\hat a_{\omega,{\bm k}_\perp}^{\rm I}\hat a_{\omega',{\bm k}_\perp'}^{\rm I \dagger}|0,{\rm M}\rangle
=\langle0,{\rm M}|\hat a_{\omega,{\bm k}_\perp}^{\rm II}\hat a_{\omega',{\bm k}_\perp'}^{\rm II \dagger}|0,{\rm M}\rangle
={1\over 1- e^{-2\pi\omega/a} }\delta_D(\omega-\omega')\delta_D^2(\bm k_\perp-\bm k_\perp'),
\nonumber
\\
&&\langle0,{\rm M}|\hat a_{\omega,{\bm k}_\perp}^{\rm I \dagger}\hat a_{\omega',{\bm k}_\perp'}^{\rm II \dagger}|0,{\rm M}\rangle
= \langle0,{\rm M}|\hat a_{\omega,{\bm k}_\perp}^{\rm I }\hat a_{\omega',{\bm k}_\perp'}^{\rm II}|0,{\rm M}\rangle
={e^{\pi\omega/a}\over e^{2\pi\omega/a}-1}\delta_D(\omega-\omega')\delta_D^2(\bm k_\perp-\bm k_\perp').
\nonumber
\end{eqnarray}
Using the definitions of the mode functions in (\ref{eqsixai}), 
we find that the two-point Wightman function becomes
\begin{eqnarray}
\langle0,{\rm M}|\phi(x)\phi(x')|0,{\rm M}\rangle
&=&\int_0^{\infty}d\omega\int_{-\infty}^{\infty} 
 d^2 k_\perp \left\{ w_{-\omega,\bm k_\perp}(x)w^*_{-\omega,\bm k_\perp}(x')
+w_{\omega,\bm k_\perp}(x)w^*_{\omega,\bm k_\perp}(x')\right\}.
\end{eqnarray}
This is equivalent to the two-point function 
\begin{eqnarray}
  \langle0,{\rm M}|\phi(x)\phi(x')|0,{\rm M}\rangle=\int {dk_zd^2 k_\perp\over (2\pi)^{3}2k_0}
{e^{-i k_0 (t-t'-i\varepsilon)+ik_z(z-z')+i\bm k_\perp\cdot(\bm x_\perp-\bm x_\perp')} },
\end{eqnarray}
as can be demonstrated using (\ref{definitionw}).
This result proves that the mode expansion in (\ref{mode-I-II}) correctly forms a complete set of wave functions
in the entire Minkowski spacetime.
\section{Two-dimensional massless field \label{Sec:entangeD=2} }
Here we study the entanglement structure of the Minkowski vacuum in
a two-dimensional massless scalar field. 
The modes correspond to the $\bm k_\perp=0$ modes, which may be neglected in the four-dimensional case. 
The analysis is much simpler 
and is well-known in the literature (see, e.g., \cite{Higuchi}), but it is instructive to show
how the entanglement structure differs from that in the four-dimensional case.

The quantized field in two dimensions is expanded as 
\begin{eqnarray}
\phi(t,z)&=&\int_{-\infty}^{\infty} {dk\over \sqrt{4\pi|k|}}
\left( {\hat b}_{k}e^{-i|k|t+ikz}+{\rm h.c.}  
\right).
\label{phitz2}
\end{eqnarray}
The Minkowski vacuum state $|0,{\rm M}\rangle$ is defined by 
$\hat b_{k}|0,{\rm M}\rangle=0, $
for~any~$k$.

\begin{figure}[t]
\begin{center}
    \includegraphics[width=7.5cm]{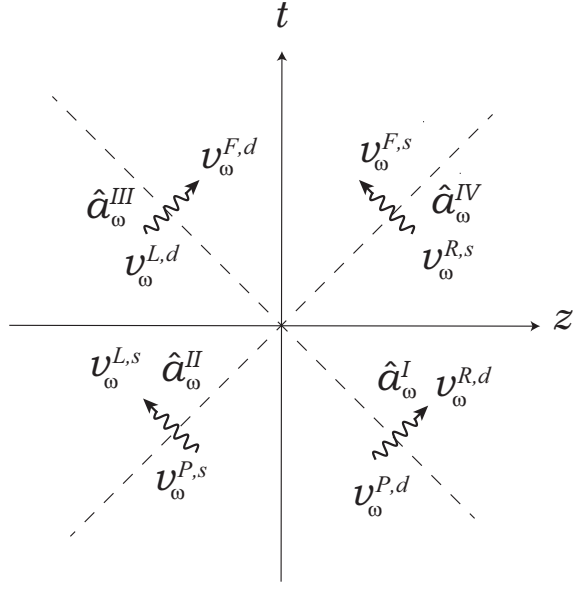}
\caption{Behavior of mode functions in the two-dimensional case.
\label{fig:coordinate2}}
\end{center}
\end{figure}

We now describe the Minkowski vacuum state 
using the quantum states constructed in the R, L, F, and 
P regions.  
For a massless scalar field in two dimensions, 
owing to conformal invariance, 
the solution for the mode function is written in a form similar to that of
the Minkowski coordinates. Then, we separate
the solutions in each region into the right-moving and left-moving waves, 
for which we use notation similar to that for the four-dimensional case. 
We define the mode functions as follows: 
\begin{eqnarray}
&&v_{\omega}^{\rm I}(x)= \theta(-U) {1\over {\sqrt{4\pi \omega}}} (-aU)^{i\omega/a}=
\left\{
\begin{array}{lc}
0 & ~~{\rm F}
\\
v^{\R,\d}_{\omega}={e^{-i\omega(\tau-\xi)}/ \sqrt{4\pi \omega}}& ~~{\rm R}
\\
0& ~~{\rm L}
\\
v_{\omega}^{\P,\d}={e^{-i\omega(\tilde\eta-\tilde\zeta)}/ \sqrt{4\pi \omega}} & ~~{\rm P}
\end{array}
\right.,
\nonumber
\end{eqnarray}
\begin{eqnarray}
&&v_{\omega}^{\rm II}(x)=\theta (-V) {1 \over \sqrt{4\pi \omega}} (-aV)^{i{\omega/ a}}=
\left\{
\begin{array}{lc}
0& ~~{\rm F}
\\
0& ~~{\rm R}
\\
v^{\L,\s}_{\omega}={e^{-i\omega(\tilde\tau-\tilde\xi)}/ \sqrt{4\pi \omega}}& ~~{\rm L}
\\
v_{\omega}^{\P,\s}={e^{-i\omega(\tilde\eta+\tilde\zeta)}/ \sqrt{4\pi \omega}} & ~~{\rm P}
\end{array}
\right.,
\nonumber
\end{eqnarray}
\begin{eqnarray}
&&v_{\omega}^{\III}(x)=\theta (U) {1 \over \sqrt{4\pi \omega}} (aU)^{-i{\omega / a}}=
\left\{
\begin{array}{lc}
v_{\omega}^{\F,\d}={e^{-i\omega(\eta-\zeta)}/ \sqrt{4\pi \omega}} & ~~{\rm F}
\\
0& ~~{\rm R}
\\
v^{\L,\d}_{\omega}={e^{-i\omega(\tilde\tau+\tilde\xi)}/ \sqrt{4\pi \omega}}& ~~{\rm L}
\\
0& ~~{\rm P}
\end{array}
\right.,
\nonumber
\end{eqnarray}
\begin{eqnarray}
&&v_{\omega}^{\rm IV}(x)=\theta (V) {1 \over \sqrt{4\pi \omega}} (aV)^{-i{\omega / a}}=
\left\{
\begin{array}{lc}
v_{\omega}^{\F,\s}={e^{-i\omega(\eta+\zeta)}/ \sqrt{4\pi \omega}} 
 & ~~{\rm F}
\\
v^{\R,\s}_{\omega}={e^{-i\omega(\tau+\xi)}/ \sqrt{4\pi \omega}}& ~~{\rm R}
\\
0& ~~{\rm L}
\\
0& ~~{\rm P}
\end{array}
\right.,
\nonumber
\end{eqnarray}
where we introduced the light-cone coordinates $U$ and $V$ in Minkowski spacetime as 
$U=t-z$ and $V=t+z$. The modes are illustrated in Fig. \ref{fig:coordinate2}.
A notable difference from the four-dimensional case with ${\bold k}_{\perp} \neq 0$ is that
all the modes propagate in the light-cone directions and cover only two regions, namely,
P and R (or L), or F and R (or L). 

With these mode functions and the creation and annihilation operators satisfying
$[\hat a^{\rm \sigma}_{\omega},\hat a^{\rm \sigma'\dagger}_{\omega'}]=\delta_{\sigma,\sigma'}\delta_D(\omega-\omega), 
~~~
[\hat a^{\rm \sigma}_{\omega},\hat a^{\rm \sigma'}_{\omega'}]=0, 
~~~
[\hat a^{\rm \sigma\dagger}_{\omega},\hat a^{\rm \sigma'\dagger}_{\omega'}]=0,
$
where $\sigma,\sigma'={\rm I,II,III,IV}$, we can expand the quantum field as
\begin{eqnarray}
&&\phi(x)=\sum_{\sigma={\rm I,II,III,IV}}\int_0^\infty d\omega
\left(
\hat a^\sigma_{\omega}v^\sigma_{\omega}(x)+{\rm h.c.}
\right).
\end{eqnarray}
Further, the Minkowski vacuum state is given by
\begin{eqnarray}
\hspace{-5mm}
  |0,{\rm M}\rangle=\prod_\omega\Bigl[
N_\omega\sum_{n_\omega=0}^\infty e^{-\pi n_\omega \omega/a }|n_\omega,{\I}\rangle \otimes|n_\omega,{\III}\rangle
\Bigr]\otimes 
\prod_{\omega'}\Bigl[
N_{\omega'}\sum_{n_{\omega'}=0}^\infty e^{-\pi n_{\omega'} \omega'/a }|n_{\omega'},{\II}\rangle \otimes|n_{\omega'},{\IV}\rangle
\Bigr] ,
\label{MIIItwo}
\end{eqnarray}
where $N_\omega=\sqrt{1-e^{-2\pi \omega/a}}$. 
Eq.~(\ref{MIIItwo}) expresses the quantum entanglement in the Minkowski vacuum 
state of the massless field in two-dimensional Minkowski spacetime, 
which also describes the modes with $\bm k_\perp=0$ in four-dimensional Minkowski spacetime
(see Fig. \ref{fig:coordinate2}). 
In the two-dimensional massless case, owing to conformal invariance, the left- and right-moving modes
are decoupled, and so is the vacuum. In the four-dimensional case, a single wave function, $v^I$, describes
a mode propagating from the P region through the R Rindler wedge to the F region. 
In contrast, the P and F regions are not connected by a single wave function in the two-dimensional case
because the mode functions $v_\omega^{\rm I}$, $v_\omega^{\rm II}$, $v_\omega^{\rm III}$, and $v_\omega^{\rm IV}$
are functions of a single positive variable, $\pm U$ or $\pm V$.  
This is the reason for the absence of quantum radiation in the two-dimensional massless scalar studied 
in Sec.~\ref{Sec:radiation-in-d2}.


In the rest of this section,
 we compute the two-point Wightman function and show that the above mode expansion
forms a complete basis in two-dimensional Minkowski spacetime.
The two-point Wightman function is given by
\begin{eqnarray}
&&\langle0,{\rm M}|\phi(x)\phi(x')|0,{\rm M}\rangle=\int_0^\infty d\omega \biggl[\left\{
v_{\omega}^{\rm I}(x)v_{\omega}^{\rm I*}(x')
+v_{\omega}^{\rm III}(x)v_{\omega}^{\rm III*}(x')
\right\}{1\over 1-e^{-2\pi\omega/a}}
\nonumber\\
&&+\left\{
v_{\omega}^{\rm I*}(x)v_{\omega}^{\rm I}(x')
+v_{\omega}^{\rm III*}(x)v_{\omega}^{\rm III}(x')
\right\}{1\over e^{2\pi\omega/a}-1}
+\bigl\{
v_{\omega}^{\rm I}(x)v_{\omega}^{\rm III}(x')
+v_{\omega}^{\rm I*}(x)v_{\omega}^{\rm III*}(x')
\nonumber\\
&&+
v_{\omega}^{\rm III}(x)v_{\omega}^{\rm I}(x')
+v_{\omega}^{\rm III*}(x)v_{\omega}^{\rm I*}(x')
\bigr\}{e^{\pi\omega/a}\over e^{2\pi\omega/a}-1}
+\left\{v_{\omega}^{\rm II}(x)v_{\omega}^{\rm II*}(x')
+v_{\omega}^{\rm IV}(x)v_{\omega}^{\rm IV*}(x')
\right\}{1\over 1-e^{-2\pi\omega/a}}
\nonumber\\
&&+\left\{
v_{\omega}^{\rm II*}(x)v_{\omega}^{\rm II}(x')
+v_{\omega}^{\rm IV*}(x)v_{\omega}^{\rm IV}(x')
\right\}{1\over e^{2\pi\omega/a}-1}
+\bigl\{
v_{\omega}^{\rm II}(x)v_{\omega}^{\rm IV}(x')
+v_{\omega}^{\rm II*}(x)v_{\omega}^{\rm IV*}(x')
\nonumber\\
&&+
v_{\omega}^{\rm IV}(x)v_{\omega}^{\rm II}(x')
+v_{\omega}^{\rm IV*}(x)v_{\omega}^{\rm II*}(x')
\bigr\}{e^{\pi\omega/a}\over e^{2\pi\omega/a}-1}\biggr],
\end{eqnarray}
where we used
\begin{eqnarray}
&& \langle0,{\rm M}|\hat a_{\omega}^{\rm I}\hat a_{\omega'}^{\rm I \dagger}|0,{\rm M}\rangle
=\langle0,{\rm M}|\hat a_{\omega}^{\rm III}\hat a_{\omega'}^{\rm III \dagger}|0,{\rm M}\rangle
={1\over 1- e^{-2\pi\omega/a} }\delta_D(\omega-\omega'),
\nonumber
\\
&&\langle0,{\rm M}|\hat a_{\omega}^{\rm I \dagger}\hat a_{\omega'}^{\rm III \dagger}|0,{\rm M}\rangle
= \langle0,{\rm M}|\hat a_{\omega}^{\rm I }\hat a_{\omega'}^{\rm III}|0,{\rm M}\rangle
={e^{\pi\omega/a}\over e^{2\pi\omega/a}-1}\delta_D(\omega-\omega'),
\nonumber
\\
&& \langle0,{\rm M}|\hat a_{\omega}^{\rm II}\hat a_{\omega'}^{\rm II \dagger}|0,{\rm M}\rangle
=\langle0,{\rm M}|\hat a_{\omega}^{\rm IV}\hat a_{\omega'}^{\rm IV \dagger}|0,{\rm M}\rangle
={1\over 1- e^{-2\pi\omega/a} }\delta_D(\omega-\omega'),
\nonumber
\\
&&\langle0,{\rm M}|\hat a_{\omega}^{\rm II \dagger}\hat a_{\omega'}^{\rm IV \dagger}|0,{\rm M}\rangle
= \langle0,{\rm M}|\hat a_{\omega}^{\rm II }\hat a_{\omega'}^{\rm IV}|0,{\rm M}\rangle
={e^{\pi\omega/a}\over e^{2\pi\omega/a}-1}\delta_D(\omega-\omega') .
\nonumber
\\
\end{eqnarray}
The others are zero. 
In terms of the positive-frequency Minkowski  modes defined as 
\begin{eqnarray}
&&F_\omega(U)=\sqrt{1-e^{-2\pi\omega/a}}\int_0^\infty {dk\over \sqrt{4\pi k}} \alpha_{\omega k}^{\rm R} e^{-ik U},
\hspace{10mm}
\bar F_\omega(U)=\sqrt{1-e^{-2\pi\omega/a}}\int_0^\infty {dk\over \sqrt{4\pi k}} \alpha_{\omega k}^{\rm L} e^{-ik U},
\nonumber\\ 
&&G_\omega(V)=\sqrt{1-e^{-2\pi\omega/a}}\int_0^\infty {dk\over \sqrt{4\pi k}} \alpha_{\omega k}^{\rm R} e^{-ik V},
\hspace{10mm}
\bar G_\omega(V)=\sqrt{1-e^{-2\pi\omega/a}}\int_0^\infty {dk\over \sqrt{4\pi k}} \alpha_{\omega k}^{\rm L} e^{-ik V},
\nonumber
\end{eqnarray}
with
\begin{eqnarray}
&& \alpha_{\omega k}^{\rm R}={ie^{\pi\omega/2a}\over 2\pi\sqrt{\omega k}}\biggl({a\over k}\biggr)^{-i\omega/a}\Gamma(1-i\omega/a),
\hspace{10mm}
\alpha_{\omega k}^{\rm L}=-{ie^{\pi\omega/2a}\over 2\pi\sqrt{\omega k}}\biggl({a\over k}\biggr)^{i\omega/a}\Gamma(1+i\omega/a),
\nonumber
\end{eqnarray}
the wave functions $v^{\rm I, II, III, IV}$
 are written as (see, e.g.,  Ref.~\cite{Higuchi}):
\begin{eqnarray}
&&\theta(-U)v^{\rm I}_{\omega}(x)
=  \frac{\bar F_{\omega}(U) - e^{-\pi\omega/a} F_{\omega}^{*}(U)}{\sqrt{1 - e^{-2\pi\omega/a}}},
\label{eqfiveai2}
\\
&&\theta(-V)v^{\rm II}_{\omega}(x)
 =  \frac{\bar G_{\omega}(V) - e^{-\pi\omega/a}G_{\omega}^{*}(V)}{\sqrt{1-e^{-2\pi \omega/a}}},
\label{eqsixai2}
\\
&&\theta(U)v^{\rm III}_{\omega}(x)
=  \frac{F_{\omega}(U) - e^{-\pi\omega/a}\bar F_{\omega}^{*}(U)}{\sqrt{1 - e^{-2\pi\omega/a}}},
\label{eqsevenai2}
\\
&&\theta(V)v^{\rm IV}_{\omega}(x)
 =  \frac{G_{\omega}(V) - e^{-\pi\omega/a}\bar G_{\omega}^{*}(V)}{\sqrt{1-e^{-2\pi \omega/a}}} .
\label{eqnineai2}
\end{eqnarray}
By using these relations, the two-point Wightman function reduces to
\begin{eqnarray}
\langle0,{\rm M}|\phi(x)\phi(x')|0,{\rm M}\rangle
&=&\int_0^{\infty}d\omega\left\{ F_{\omega}(x)F^*_{\omega}(x')+\bar F_{\omega}(x)\bar F^*_{\omega}(x')
    +G_{\omega}(x)G^*_{\omega}(x')+\bar G_{\omega}(x)\bar G^*_{\omega}(x')
    \right\}.
    \label{FFFFGGGG}
\end{eqnarray}
By using the relation
\begin{eqnarray}
  \int_{-\infty}^\infty d\omega(1-e^{-2\pi\omega/a})(\alpha_{\omega k}^{\rm R}\alpha_{\omega k'}^{\rm R*}+
  \alpha_{\omega k}^{\rm L}\alpha_{\omega k'}^{\rm L*} )=\delta_D(k-k'), 
\end{eqnarray}
(\ref{FFFFGGGG}) becomes 
\begin{eqnarray}
  \langle0,{\rm M}|\phi(x)\phi(x')|0,{\rm M}\rangle
  =\int_0^\infty {dk\over 4\pi k}(e^{-ik(V-V')}+e^{-ik(U-U')})=
  \int_{-\infty}^\infty {dk\over 4\pi |k|}e^{-i|k|(t-t')+ik(z-z')} ,
\end{eqnarray}
which is nothing but the expression obtained directly from the Minkowski mode expansion (\ref{phitz2}).

\section{Application I: Quantum radiation from a uniformly accelerating detector in four-dimensional spacetime
\label{Sec:Application1}}
\begin{figure}[t]
\begin{center}
    \includegraphics[width=7.5cm]{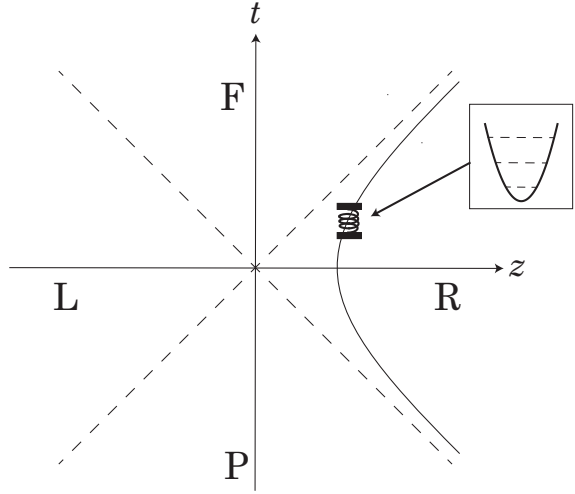}
\caption{Schematic of the Unruh--DeWitt detector model. 
The hyperbolic curve is the detector's trajectory, 
and the detector is an infinitely small harmonic oscillator coupled to the massless scalar field.
\label{fig:conceptualmodel} }
\end{center}
\end{figure}

As an application of the entanglement structure investigated so far, we study the
quantum radiation emanating from 
an Unruh--DeWitt detector coupled to a massless scalar field in four-dimensional spacetime \cite{LH,LH2,IYZ,IYZ2013,IOTYZ,ITUY}. 
The radiation is caused by the two-point correlations between the R region and
the F region based on (\ref{MIII}), which cannot be calculated
using the ordinary expression (\ref{RLentangle}). 
A two-dimensional case is studied in the next section. 

The Unruh--DeWitt detector model contains a harmonic oscillator $Q(\tau)$ moving at uniform acceleration 
with the world-line trajectory $z(\tau)$, which is coupled to the
massless scalar field $\phi$ in four-dimensional spacetime  
(see Fig. \ref{fig:conceptualmodel}).
The action is given by
\begin{eqnarray}  
S[Q,\phi ; z] &=&
\frac{m}{2} \int d \tau \left( \dot{Q}^2(\tau) - \Omega_0^2 Q^2(\tau) \right) 
+ {1\over 2}\int d^4 x \partial^\mu \phi(x) \partial_\mu \phi(x)  
\nonumber \\
&&+ \lambda \int d^4 x d\tau P[Q(\tau)] \phi(x) \delta^{(4)}_D(x-z(\tau)),
\label{actionUD}
\end{eqnarray} 
where $m$ and $\Omega_0$ are the mass and angular frequency 
of the harmonic oscillator $Q$, respectively.
The dot denotes differentiation with respect to $\tau$.
Here $P[Q]$ is defined as 
\begin{eqnarray}
P[Q(\tau)]=\sum_jp_j{d^j Q(\tau)\over d\tau^j},
\label{devPQ}
\end{eqnarray}
where $p_i$ are constants \cite{IYZ2013}. 

The world-line trajectory of the detector is specified by
$x^\mu=z^\mu(\tau)$, where $\tau$ is the proper time of the detector. 
The trajectory under uniformly accelerated motion is given by 
\begin{eqnarray}
t(\tau)=a^{-1}\sinh a\tau, ~~z(\tau)=a^{-1}\cosh a\tau, ~~\bm x_\perp(\tau)=0.
\label{trajec}
\end{eqnarray} 
The equations of motion for $Q(\tau)$ and $\phi(x)$ are given by 
\begin{eqnarray}
&&\ddot Q(\tau)+\Omega_0^2Q(\tau)={\lambda\over m}\bar P[\phi(z(\tau))]
\label{eqQ}, \\
&&\partial^\mu \partial_\mu\phi(x)=\lambda\int d\tau P[Q(\tau)]\delta^{(4)}_D(x-z(\tau)),
\label{eqphiA}
\end{eqnarray}
where we defined $\bar P[\phi(z(\tau))]=\sum_jp_j(-1)^j{d^j \phi(z(\tau))\over d\tau^j}$.
These are linearly coupled equations; hence, the system
can be solved exactly~\cite{LH,LH2}. 
As we show in the following, the equation of motion (\ref{eqQ}) becomes a Langevin-type equation, 
and the harmonic oscillator, after a transient phase, 
 eventually becomes thermalized to an equilibrium state at the Unruh temperature,
$T_U=a/2\pi$. Consequently, the scalar field $\phi$ is also stabilized to a steady state.
We thus consider such an equilibrium phase in the following investigation \cite{IYZ,IYZ2013}.
In the presence of the harmonic oscillator, 
the scalar field $\phi(x)$ is given by the sum of the homogeneous solution $\phi_{\rm h}(x)$ and the 
inhomogeneous solution $\phi_{\rm inh}(x)$: 
\begin{eqnarray}
  \phi(x)=\phi_{\rm h}(x)+\phi_{\rm inh}(x) .
\label{phpinh}
\end{eqnarray}
The homogeneous solution $\phi_{\rm h}(x)$ represents the vacuum fluctuation and always exists independent of
the presence or absence of $Q(\tau)$.
In contrast, the inhomogeneous solution 
 $\phi_{\rm inh}(x)$ is generated by the harmonic oscillator $Q(\tau)$ and is given by
\begin{eqnarray}
\phi_{\rm inh}(x) =\lambda \int d\tau' P[Q(\tau')]G_R(x-z(\tau')), 
\label{phiQG}
\end{eqnarray}
where $G_R(x-y)$ is the retarded Green function. 
Note that the harmonic oscillator $Q(\tau)$ is considered to be in the classical ground state 
and to be excited only through interaction with the quantum fluctuations of the scalar field 
$\phi_{\rm h}(x)$, as is shown below. 
Thus, $Q(\tau)$ is determined by the quantum field $\phi_{\rm h}(z(\tau))$ 
on the trajectory of the detector 
$z(\tau)$. 

By inserting (\ref{phpinh}) into (\ref{eqQ}), we obtain
\begin{eqnarray}
\ddot Q+\Omega^2_0 Q-{\lambda^2\over m}\bar P\left[
\int d\tau' P[Q(\tau')]G_R(z(\tau)-z(\tau'))\right]
={\lambda\over m}\bar P[\phi_{\rm h}(z(\tau))]. 
\label{eqQQ}
\end{eqnarray}
In the four-dimensional massless case, the retarded Green function is given by
\begin{eqnarray}
G_R(x-y)={1\over 4\pi}\delta_D(\sigma^2(x-y))\theta(x^0-y^0)
, ~~~\sigma^2(x-y)={1\over 2}(x_\mu-y_\mu)(x^\mu-y{}^\mu).
\end{eqnarray}
To regularize the ultraviolet divergences in the $\tau'$ integral at $\tau'=\tau$,
we introduce the regularized retarded Green function \cite{LH},
\begin{eqnarray}
G_R^\Lambda(x-x')={1\over 4}\sqrt{8\over\pi}\Lambda^2 e^{-2\Lambda^4\sigma^2(x-x')}\theta(x^0-x'{}^0),
\label{rrgf}
\end{eqnarray}
where $\Lambda$ is the regularization parameter. 
Then, we have
\begin{eqnarray}
\phi_{\rm inh}(z(\tau))&=&\lambda \int d\tau' P[Q(\tau')] G_R^\Lambda(z(\tau)-z(\tau'))
\nonumber\\
&=&{\lambda\over 4\pi}\left\{\Lambda\zeta P[Q(\tau)]-{d\over d\tau}P[Q(\tau)]+{\cal O}(\Lambda^{-1})\right\},
\label{rrgf22}
\end{eqnarray}
where $\zeta=2^{7/4}\Gamma(5/4)/\sqrt{\pi}$. In the large-$\Lambda$ limit, the ${\cal O}(\Lambda^{-1})$ terms
can be dropped.

When $P[Q]=Q$, (\ref{eqQQ}) is simplified as follows. 
By inserting the solution (\ref{rrgf22}) into (\ref{eqQQ}),
we find that $Q(\tau)$ satisfies the equation of motion
\begin{eqnarray}
\ddot Q+2\gamma \dot Q +\Omega^2 Q={\lambda\over m}\phi_{\rm h}(z(\tau)),
\end{eqnarray}
where we introduced $\gamma=\lambda^2/8\pi m$ and the renormalized frequency $\Omega^2
=\Omega_0^2-\lambda^2\Lambda\zeta/4\pi m$.
This is the Langevin equation with the dissipation term coming from the radiation reaction term,
whereas the noise term on the right-hand side comes from the quantum fluctuation $\phi_{\rm h}(z(\tau))$
on the trajectory $z(\tau)$. 
Using the Fourier-transformed variables 
\begin{eqnarray}
&&Q(\tau)={1\over2\pi}\int_{-\infty}^{\infty} d\omega e^{-i\omega\tau}\tilde Q(\omega),
\label{solQtau}
\\
&&\phi_{\rm h}(z(\tau))={1\over2\pi}\int_{-\infty}^{\infty} d\omega e^{-i\omega\tau}\varphi(\omega),
\label{solphih}
\end{eqnarray}
we find the solution in the equilibrium state:
\begin{eqnarray}
&&\tilde Q(\omega)=\lambda h(\omega) \varphi(\omega) , \hspace{5mm}
h(\omega)={1\over -m\omega^2+m\Omega^2-i2m{\omega\gamma}}. 
\label{solQ}
\end{eqnarray}

For the general case $P[Q]\neq Q$, the function $h(\omega)$ is replaced by
\begin{eqnarray}
&&h(\omega)={f(-\omega)\over -m\omega^2+m\Omega_0^2-\lambda^2f(\omega)f(-\omega)\tilde G_R(\omega)}, 
\label{defhomega}
\end{eqnarray}
where we defined $f(\omega)=\sum_j p_j(-i\omega)^j$. The retarded Green function satisfies
\begin{eqnarray}
&&\tilde G_R(\omega)= \int d(\tau-\tau') G_R(\tau-\tau') e^{i\omega(\tau-\tau')}=\tilde G_R^*(-\omega);
\label{deftildeGR}
\end{eqnarray}
hence, the relation $h(\omega)=h^*(-\omega)$ holds. 

Thus, from (\ref{phiQG}), we have the expression for the inhomogeneous solution: 
\begin{eqnarray}
\phi_{\rm inh}(x)&=&\lambda^2 \int d\tau\int {d\omega\over2\pi}e^{-i\omega\tau}
f(\omega)h(\omega)G_R(x-z(\tau))\varphi(\omega)
\nonumber\\
&=&{\lambda^2 \over 4\pi \rho_0(x)}\int {d\omega\over2\pi}e^{-i\omega\tau_-^x}
f(\omega)h(\omega)\varphi(\omega),
\label{phiinh2}
\end{eqnarray}
where $\rho_0(x)$ is defined as 
\begin{eqnarray}
\rho_0(x)={a\over 2}\sqrt{\left(L^2\right)^2+{4\over a^2}(t^2-z^2)}
\label{defrho0}
\end{eqnarray}
with
\begin{eqnarray}
L^2=-x^\mu x_\mu+{1\over a^2}=-t^2+z^2+\bm x_\perp^2+{1\over a^2}. 
\end{eqnarray}
The factor $\rho_0(x)$ in the second equality of (\ref{phiinh2})  
comes from the Jacobian used to evaluate $\delta(\sigma^2(x-z(\tau)))$ 
in the $\tau$ integration, and $\tau_-^x$ is the proper time on the detector's trajectory at which
a past light cone from position $x$ intersects it (see Fig. \ref{fig:taupm}), 
as defined in (\ref{deftaum}).
Note that the inhomogeneous solution $\phi_{\rm inh}(x)$ is determined by 
the quantum fluctuation $\varphi(\omega)$, that is, the Fourier transform of 
 $\phi_{\rm h}(z(\tau))$ on the trajectory in the R Rindler wedge.

We next investigate the behavior of  the two-point correlation function, 
\begin{eqnarray}
\langle\phi_{\rm }(x)\phi_{\rm }(y)\rangle=\langle\phi_{\rm h}(x)\phi_{\rm h}(y)\rangle+
\langle\phi_{\rm h}(x)\phi_{\rm inh}(y)\rangle
+\langle\phi_{\rm inh}(x)\phi_{\rm h}(y)\rangle
+\langle\phi_{\rm inh}(x)\phi_{\rm inh}(y)\rangle,
\label{pppppp}
\end{eqnarray}
from which one can calculate the energy--momentum tensor
and study the properties of the quantum radiation. 
The correlation function in (\ref{pppppp}) contains essentially three terms.
The first term, $\langle\phi_{\rm h}(x)\phi_{\rm h}(y)\rangle$, represents 
the vacuum fluctuation and always exists irrespective of the presence of $Q(\tau)$.
Thus, it is irrelevant in the present discussion and can be ignored.
The last term, $\langle\phi_{\rm inh}(x)\phi_{\rm inh}(y)\rangle$, is the {\it naive} radiation 
term. When the source term $Q(\tau)$ behaves classically, for example, for a charged particle,
it is the only term that contributes to the radiation. Indeed, when we calculate the Larmor radiation
in classical electromagnetic theory, this term gives the radiation. 
Thus, we call it the {\it naive} radiation term. In the present case, because the harmonic oscillator
is excited by the quantum fluctuation $\phi_{\rm h}(z(\tau))$, it is not sufficient to consider
only this term; it is also necessary to consider the quantum interference.
The remaining term, $\langle\phi_{\rm h}(x)\phi_{\rm inh}(y)\rangle
+\langle\phi_{\rm inh}(x)\phi_{\rm h}(y)\rangle$, represents this interference.
Thus, the issue of the quantum Unruh radiation is strongly affected by the structure of the
interference terms.

\begin{figure}[t]
\begin{center}
    \includegraphics[width=5.5cm]{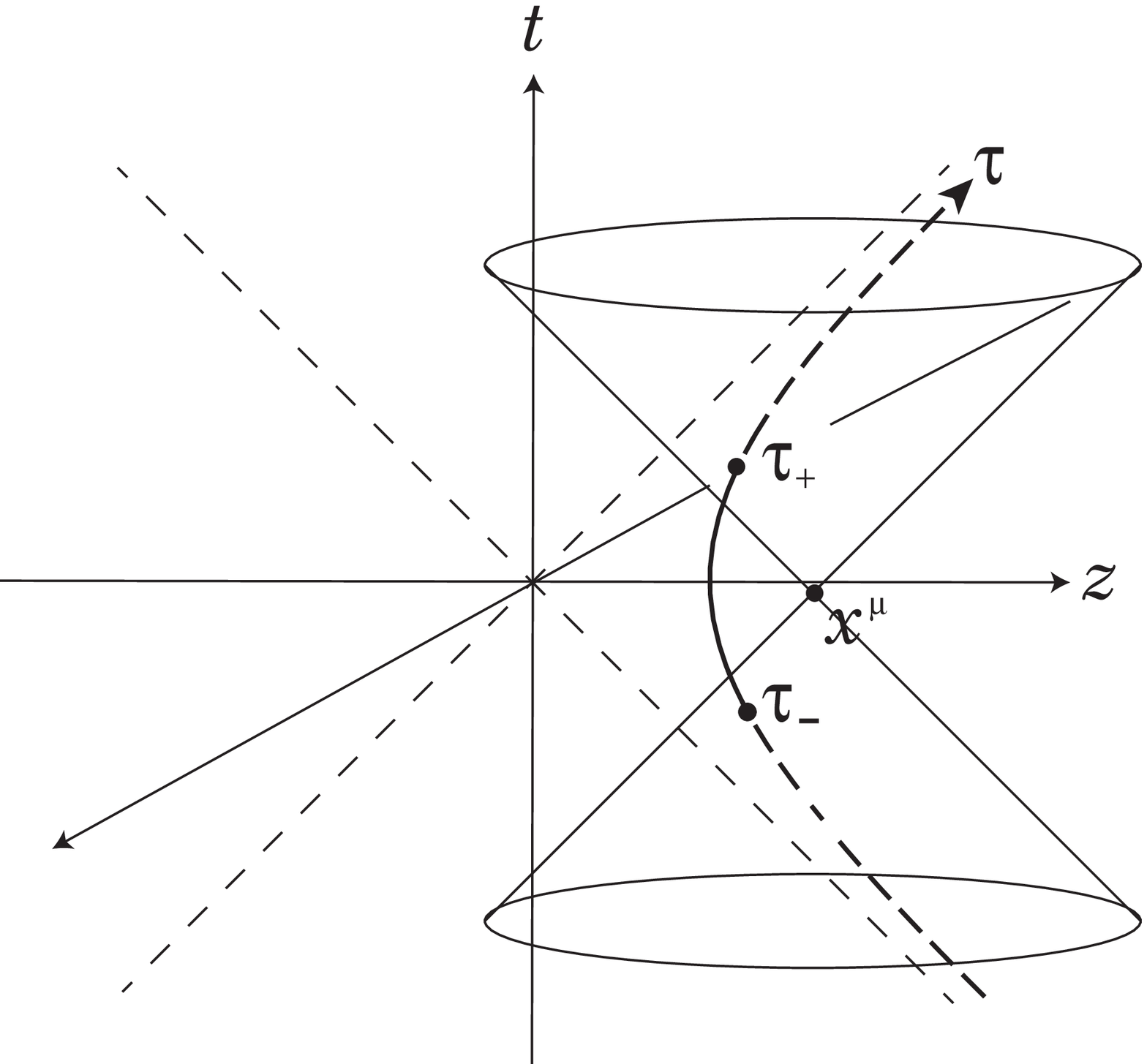}
~~~~~~~~~~~~~~~~~~~~~~
    \includegraphics[width=5.5cm]{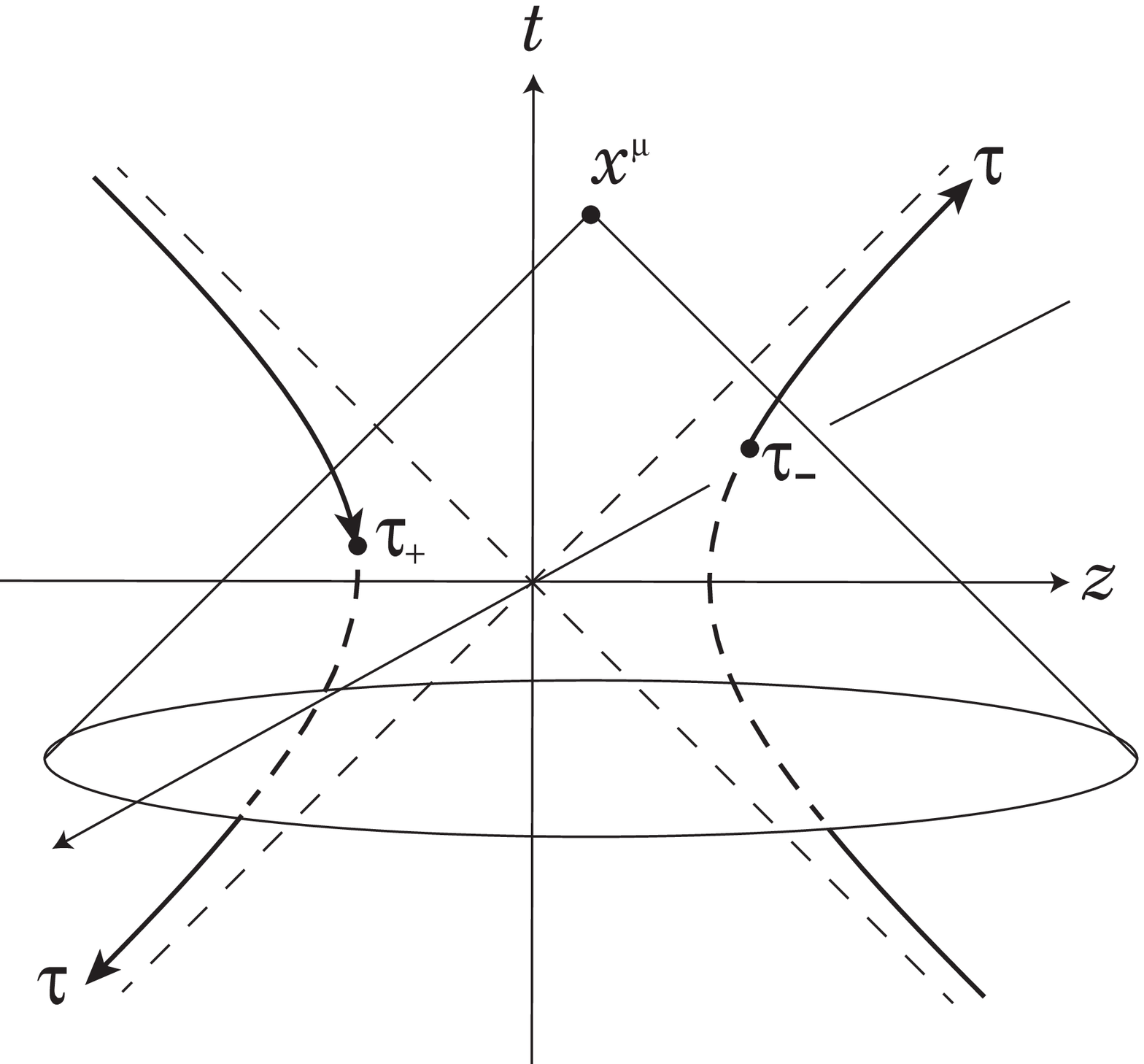}
\caption{$\tau_-^x$ is the
proper time $\tau$ of the point where the past light cone of $x^\mu$ intersects 
the detector's trajectory. 
When $x^\mu$ is in the R region, $\tau_+^x$ is the proper time $\tau$ of the 
point where the future light cone of 
$x^\mu$ intersects the detector's trajectory (left panel). 
When $x^\mu$ is in the F region, $\tau_+^x$ is the proper time $\tau$ of the 
point where the past light cone of 
$x^\mu$ intersects the hypothetical detector's trajectory in the L region, and the trajectory is
a mirror image of the real trajectory (right panel). 
\label{fig:taupm}
}
\end{center}
\end{figure}

Now we calculate the two-point functions \cite{IYZ,IYZ2013,IOTYZ}.
Since the inhomogeneous term is written in terms of the homogeneous term
 (quantum flcuatuation) on the trajectory as in Eq.~(\ref{phiinh2}). 
This means that the field $\phi_{\rm inh}(x)$ is a quantum field, and the quantity 
$\langle\phi_{\rm inh}(x)\phi_{\rm inh}(y)\rangle$ is also the vacuum expectation 
value of the product of the fields $\phi_{\rm inh}(x)$ and $\phi_{\rm inh}(y)$.
By using (\ref{phiinh2}), we straightforwardly obtain the naive radiation term as
\begin{eqnarray}
&&\langle\phi_{\rm inh}(x)\phi_{\rm inh}(y)\rangle
  ={\lambda^4\over (4\pi)^2\rho_0(x)\rho_0(y)}
  \int _{-\infty}^{\infty}{d\omega\over2\pi}  \int _{-\infty}^{\infty}{d\omega'\over2\pi}
f(\omega)h(\omega)f(\omega')h(\omega')e^{-i\omega\tau_-^x-i\omega'\tau_-^y} \langle
\varphi(\omega)\varphi(\omega')\rangle.
\label{inhomogeneoustermb}
\end{eqnarray}
With the use of (\ref{solphih}), we have
\begin{eqnarray}
\langle\varphi(\omega)\varphi(\omega')\rangle
&=&\int _{-\infty}^{\infty} d\tau \int _{-\infty}^{\infty} d\tau'\langle \phi_h(z(\tau))\phi_h(z(\tau'))\rangle
e^{i\omega\tau+i\omega'\tau'}
\nonumber\\
&=&\int _{-\infty}^{\infty} d\tau \int _{-\infty}^{\infty} d\tau'{-e^{i\omega\tau+i\omega'\tau'}\over 4\pi^2
[(t(\tau)-t(\tau')-i\epsilon)^2-(z(\tau)-z(\tau'))^2]}
\nonumber\\
&=&\int _{-\infty}^{\infty} d\tau \int _{-\infty}^{\infty} d\tau'{-e^{i\omega\tau+i\omega'\tau'}\over (4\pi)^2
\sinh^2\left\{(\tau-\tau'-i\epsilon)/2\right\}}
\nonumber\\
&=&\delta_D(\omega+\omega'){\omega\over 1-e^{-2\pi\omega/a}}, 
\end{eqnarray}
where we used the Wightman function of the massless scalar field in the four dimensions
\cite{BD}. Using Eq.~(\ref{defhomega}) and ${\rm Im}\tilde G_R(\omega)=\omega/4\pi$ (see, e.g., 
\cite{IYZ2013}), we have
\begin{eqnarray}
h(\omega)f(\omega)-h(-\omega)f(-\omega)
=2i\lambda^2 {\rm Im}\tilde G_R(\omega)|f(\omega)h(\omega)|^2 
=i\lambda^2{\omega\over 2\pi}|f(\omega)h(\omega)|^2.
\end{eqnarray}
Then, Eq.~(\ref{inhomogeneoustermb}) is written as
\begin{eqnarray}
&&\langle\phi_{\rm inh}(x)\phi_{\rm inh}(y)\rangle
  ={-i\lambda^2\over (4\pi)^2\rho_0(x)\rho_0(y)}
  \int _{-\infty}^{\infty}{d\omega\over2\pi}{1\over e^{2\pi\omega/a}-1}
\bigr[f(\omega)h(\omega)-f(-\omega)h(-\omega)\bigl]e^{i\omega(\tau_-^x-\tau_-^y)} .
\label{inhomogeneousterm}
\end{eqnarray}
Here we changed the integration variable from $\omega$ to $-\omega$.
This term gives rise to the radiation naively expected from the detector
in the thermally excited state.

On the other hand, the interference term is obtained by using  (\ref{solphih}) and
 (\ref{phiinh2}) as (see Ref.~\cite{IYZ2013}) 
\begin{eqnarray}
&&
\langle\phi_{\rm h}(x)\phi_{\rm inh}(y)\rangle
+\langle\phi_{\rm inh}(x)\phi_{\rm h}(y)\rangle
\nonumber\\
&
  &~~~~= {-i\lambda^2\over (4\pi)^2\rho_0(x)\rho_0(y)}
  \int _{-\infty}^{\infty}{d\omega\over2\pi}{1
\over e^{2\pi\omega/a}-1}
\bigr[f(\omega)h(\omega)e^{i\omega(\tau_+^x-\tau_-^y)}Z_x(\omega)
-f(-\omega)h(-\omega)e^{i\omega(\tau_-^x-\tau_+^y)}Z_y(\omega)\bigl]
\nonumber
\\
&&~~~~+{i\lambda^2\over (4\pi)^2\rho_0(x)\rho_0(y)}
  \int _{-\infty}^{\infty}{d\omega\over2\pi}{1\over e^{2\pi\omega/a}-1}
\bigr[f(\omega)h(\omega)-f(-\omega)h(-\omega)\bigl]e^{i\omega(\tau_-^x-\tau_-^y)} .
\label{interferenceall}
\end{eqnarray}
The details of the similar calculations based on the Green function method in the 
reference frame coordinates
are given in Sec.~\ref{Sec:InterferenceMink}, and calculations based on the operator formalism
with (\ref{MIII}) are given in Sec.~\ref{Sec:interference-operator}. 

In the above formula (\ref{interferenceall}), we defined 
\begin{eqnarray}
Z_x(\omega)=e^{\pi\omega/a}\theta(t-z)+\theta(-t+z),
\label{defZx}
\end{eqnarray}
and $\tau_-^x$ is given by 
\begin{eqnarray}
\tau_-={{1\over a}\log\left[{a\over 2(t-z)}\left(-L^2+\sqrt{L^4+{4\over a^2}[t^2-z^2]}
\right)\right]}
\label{deftaum}
\end{eqnarray}
for $x^\mu$ in either the R or F region. 
On the other hand, $\tau_+^x$ is given by a different solution:
\begin{eqnarray}
\tau_+=
{{1\over a}\log\left[{a\over 2(t-z)}\left(\mp L^2 \mp \sqrt{L^4+{4\over a^2}[t^2-z^2]}
\right)\right]}, 
\label{deftaup}
\end{eqnarray}
for $x^\mu$ in the R region (upper sign) or F region (lower sign). 
When $x^\mu$ is in the R region, $\tau_-^x$ ($\tau_+^x$) is the proper time of the 
detector at the point where the detector's trajectory intersects the 
past light cone (future light cone) of $x^\mu$ (see the left-hand panel of Fig.~\ref{fig:taupm}). 
On the other hand, when $x^\mu$ is in the F region, $\tau_-^x$ is 
the proper time of the detector at which the detector's trajectory intersects 
the past light cone of $x^\mu$ as above, but
$\tau_+^x$ is 
the proper time at the intersection point of the past light cone
of $x^\mu$ and the
hypothetical detector in the L region (see the right-hand panel of Fig.~\ref{fig:taupm}),
whose trajectory is defined as 
\begin{eqnarray}
t(\tau)=-a^{-1}\sinh a\tau, \  ~z(\tau)=-a^{-1}\cosh a\tau, \  ~\bm x_\perp(\tau)=0.
\label{trajecL}
\end{eqnarray}

Looking at (\ref{inhomogeneousterm}) and 
(\ref{interferenceall}),
we see that the naive radiation term $\langle\phi_{\rm inh}(x)\phi_{\rm inh}(y)\rangle$
is completely canceled by
the second part of the interference term (\ref{interferenceall}).
Therefore, we finally find 
\begin{eqnarray}
  &&\langle\phi(x)\phi(y)\rangle-\langle\phi_{\rm h}(x)\phi_{\rm h}(y)\rangle
\nonumber\\
  &&~~~~~~~~~~
={-i\lambda^2\over (4\pi)^2\rho_0(x)\rho_0(y)}
  \int _{-\infty}^{\infty}{d\omega\over2\pi}{1\over e^{2\pi\omega/a}-1}
\bigr[f(\omega)h(\omega)e^{i\omega(\tau_+^x-\tau_-^y)}Z_x(\omega)
-f(-\omega)h(-\omega)e^{i\omega(\tau_-^x-\tau_+^y)}Z_y(\omega)\bigl]
\nonumber\\
\label{interferencew}
\end{eqnarray}
for $x,y$ in the F or R region. 
From the two-point function, one can calculate the energy--momentum tensor:
\begin{eqnarray}
\hspace{-4mm}
T_{\mu\nu}
=\lim_{y\rightarrow x}\left({\partial\over \partial x^\mu }{\partial\over \partial y^\nu }-{1\over 2}
\eta_{\mu\nu}\eta^{\alpha\beta}{\partial\over \partial x^\alpha }{\partial\over \partial y^\beta }\right)
[\langle\phi(x)\phi(y)\rangle-\langle\phi_{\rm h}(x)\phi_{\rm h}(y)\rangle]_S,
\end{eqnarray} 
where the subscript $S$ indicates symmetrization over $x$ and $y$. 
The energy flux at a large distance is derived from the energy--momentum tensor, 
and the behavior is investigated in Ref.~\cite{IOTYZ}. 
The energy radiation rate is roughly estimated as $dE/dt=a^3\lambda^2/8\pi^2m\Omega^2$. 
The result is consistent with that found in Refs. \cite{LH,LH2}.

Fig.~\ref{fig:funF} shows an example of the angular distribution of the energy flux, 
which is estimated on the future light cone of the point on the detector's 
trajectory $\tau=0$, where the parameters are $\Omega/a=0.2$ and $\gamma/a=1$.  
The behavior of the energy flux depends on the parameters of the model.

\begin{figure}[t]
\begin{center}
~~~~~~~~~~~~~~~~~
    \includegraphics[width=5.cm]{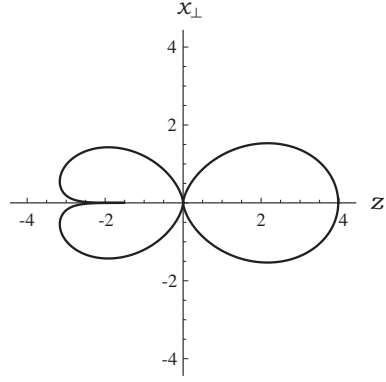}
\caption{Typical behavior of the angular distribution of the radiation flux. In general, the behavior 
of the radiation flux depends on the parameters of the model. In this case ($\Omega/a=0.2$ and $\gamma/a=1$), 
the radiation is maximum in the direction of acceleration. The position $x$ is chosen so that $\tau_-^x = 0$. 
\label{fig:funF}}
\end{center}
\end{figure}

\subsection{Calculation of $\langle\phi_{\rm h}(x)\phi_{\rm h}(z(\tau))\rangle$ with 
Green function in the reference frame coordinates 
\label{Sec:InterferenceMink}}
In this and the following sections, we calculate the 
interference term (\ref{interferenceall}) by two different methods.
One is the Green function method adopted in \cite{IYZ,IYZ2013} and reviewed in this section.
The other is the operator method based on the entanglement structure of the 
Minkowski vacuum (\ref{MIII}), which is given in the next section. This gives a physical interpretation
of the quantum radiation in the F region.

Remember that the inhomogeneous solution $\phi_{\rm inh}(x)$ is 
determined by the vacuum fluctuation $\phi_{\rm h}(z(\tau_-^x))$
on the trajectory of the accelerated motion. 
Thus, the interference term $\langle \phi_{\rm h}(x) \phi_{\rm inh}(y)\rangle$ in (\ref{interferencew})
is essentially given by the two-point correlation function  
$\langle \phi_{\rm h}(x) \phi_{\rm h}(z(\tau_-^y))\rangle$
of the vacuum fluctuations, 
one component of which, $\phi_{\rm h}(x)$, is in the F region for $x \in {\rm F}$, and the other, $\phi_{\rm h}(z(\tau_-^y))$,
is in the R region, $z(\tau_-^y) \in {\rm R}$. 
This is the key quantity for clarifying the origin of the quantum radiation, which we examine in the 
following sections.

We briefly review the calculation of this quantity using the Green function 
method in the reference frame coordinates 
 \cite{IYZ,IYZ2013}. 
The Fourier transform of $\langle\phi_{\rm h}(x)\phi_{\rm h}(z(\tau))\rangle$ becomes
\begin{eqnarray}
\langle\phi_{\rm h}(x)\varphi(\omega)\rangle
=\int d\tau e^{i\omega\tau}\langle\phi_{\rm h}(x)\phi_{\rm h}(z(\tau))\rangle =-{1\over4\pi^2} P(x,\omega),
\end{eqnarray}
where we defined 
\begin{eqnarray}
P(x,\omega)=\int d\tau {e^{i\omega\tau}\over (t-z^0(\tau)-i\epsilon)^2-(x-z^1(\tau))^2-x_\perp^2}.
\label{Pxomegaintegration}
\end{eqnarray}
The poles of the denominator of (\ref{Pxomegaintegration}) are obtained by solving
$(t-z^0(\tau)-i\epsilon)^2-(x-z^1(\tau))^2-x_\perp^2=0$, which yields
\begin{eqnarray}
\tau=
\left\{
\begin{array}{lc}
\displaystyle{
\tau_\pm-i\epsilon+i2\pi na}
~~~~&~~~~x{\rm ~in~R~region}
\\
\displaystyle{\tau_--i\epsilon+i2\pi na,~\tau_++i\pi a+i2\pi na}
~~~~&~~~~x{\rm ~in~F~region}
\end{array}
\right.
\end{eqnarray}
where $n$ takes integer values, $n=0,\pm1,\pm2\cdots$. 
The integration in (\ref{Pxomegaintegration}) yields
\begin{eqnarray}
P(x,\omega)=
\displaystyle{{\pi i\over \rho_0}\left({Z_x(\omega)\over e^{2\pi\omega/a}-1}e^{i\omega\tau_+^x}
-{1\over e^{2\pi\omega/a}-1}e^{i\omega\tau_-^x}\right)}
\end{eqnarray}
for $x$ in the R or F region, and we obtain \cite{IYZ,IYZ2013}
\begin{eqnarray}
\langle\phi_{\rm h}(x)\phi_{\rm h}(z(\tau))\rangle=
-{i\over 8\pi^2\rho_0(x)}
\displaystyle{
\int_{-\infty}^{\infty}d\omega e^{-i\omega\tau}
\left(
{Z_x(\omega)\over e^{2\pi\omega/a}-1}e^{i\omega\tau_+^x}
-{1\over e^{2\pi\omega/a}-1}e^{i\omega\tau_-^x}
\right)}.
\label{resultoffs}
\end{eqnarray}

The first and second terms on the right-hand side of (\ref{resultoffs})
correspond to the first and second terms on the right-hand side of 
(\ref{interferenceall}).\footnote{Note that the interference term in (\ref{interferenceall}) 
represents a combination of $\langle\phi_{\rm h}(x)\phi_{\rm inh}(y)\rangle
+\langle\phi_{\rm inh}(x)\phi_{\rm h}(y)\rangle$. }
As we saw, 
only the first term in (\ref{interferenceall}) contributes to 
the quantum radiation, whereas the second term in (\ref{interferenceall}) cancels out the naive radiation term. 
Because the physical origins of these two terms are not clear in the Green function 
method in the reference frame coordinates,
we will re-derive it in the next section 
using the operator formalism based on the discussions in Sec. \ref{Sec:entanglement}.

\subsection{Calculation of $\langle\phi_{\rm h}(x)\phi_{\rm h}(z(\tau))\rangle$ in the operator 
formalism with (\ref{MIII})
\label{Sec:interference-operator}
}
We compute the correlation function 
$\langle0,{\rm M}|\phi_{\rm h}(x) \phi_{\rm h}(z(\tau))|0,{\rm M}\rangle$  
using the operator formalism on the Minkowski vacuum (\ref{MIII}). 
Here we focus on $x$ in the F region, but 
we can show that a similar result is obtained when $x$ is in the R region (see Appendix B). 

When $x$ is in the F region, the correlation function becomes
\begin{eqnarray}
&&\langle0,{\rm M}| \phi_{\rm h}(x) \phi_{\rm h}(z(\tau))|0,{\rm M}\rangle=
\langle0,{\rm M}| \phi_{\rm h}^{\rm F,d}(x) \phi_{\rm h}(z(\tau))|0,{\rm M}\rangle+
\langle0,{\rm M}| \phi_{\rm h}^{\rm F,s}(x) \phi_{\rm h}(z(\tau))|0,{\rm M}\rangle,
\end{eqnarray}
where 
\begin{eqnarray}
&&\langle0,{\rm M}| \phi_{\rm h}^{\rm F,d}(x) \phi_{\rm h}(z(\tau))|0,{\rm M}\rangle
\nonumber\\
&&~~~~~~=\int_0^{\infty}d\omega\int\int d^2k_\perp\biggl(
v_{\omega,\bm k_\perp}^{\rm F,d}(x)v_{\omega,\bm k_\perp}^{\rm R}(z(\tau))
{e^{\pi\omega/a}\over e^{2\pi\omega/a}-1}
+v_{\omega,\bm k_\perp}^{\rm F,d*}(x)v_{\omega,\bm k_\perp}^{\rm R*}(z(\tau))
{e^{\pi\omega/a}\over e^{2\pi\omega/a}-1}\biggr),
\label{MppM1}
\\
&&\langle0,{\rm M}| \phi_{\rm h}^{\rm F,s}(x) \phi_{\rm h}(z(\tau))|0,{\rm M}\rangle
\nonumber\\
&&~~~~~=
\int_0^{\infty}d\omega\int\int d^2k_\perp\biggl(
v_{\omega,\bm k_\perp}^{\rm  F,s*}(x)v_{\omega,\bm k_\perp}^{\rm R}(z(\tau))
{1\over e^{2\pi\omega/a}-1}
+v_{\omega,\bm k_\perp}^{\rm F,s}(x)v_{\omega,\bm k_\perp}^{\rm R*}(z(\tau))
{1\over 1-e^{-2\pi\omega/a}}
\biggr) .
\label{MppM2}
\end{eqnarray}
Here we used the expression of the Minkowski vacuum in (\ref{MIII}) and 
the results in Sec.~\ref{Sec:2pointWightman}.

The coordinates of $x$ in the F region are specified as $(\eta,\zeta,\bm x_\perp)$.
We first focus on the first term of the integration in (\ref{MppM1}). It becomes
\begin{eqnarray}
&&\int_0^{\infty}d\omega\int\int d^2k_\perp
v_{\omega,\bm k_\perp}^{\rm F,d}(x)v_{\omega,\bm k_\perp}^{\rm R}(z(\tau))
{e^{\pi\omega/a}\over e^{2\pi\omega/a}-1}
\nonumber\\
&&={-i\over \sqrt{16 a^2\pi^4}}\int_0^{\infty}d\omega e^{i\omega\zeta-i\omega\tau}
{e^{\pi\omega/a}\over e^{2\pi\omega/a}-1} \int_0^\infty d\kappa \kappa 
J_{-i\omega/a}\left({\kappa\over a}e^{a\eta}\right)
K_{-i\omega/a}\left({\kappa\over a}\right) J_0(\kappa x_\perp),
\label{firsttermMppM}
\end{eqnarray}
where we used the expression
$J_0(x)={1\over 2\pi} \int_0^{2\pi} d\varphi e^{ix\cos\varphi}$
and the relation
$ I_\nu(z)=e^{-\nu\pi i/2}J_\nu(e^{\pi i/2}z) $ 
which holds for $-\pi<\arg z<\pi/2$. 
Then, by using the mathematical formula
\begin{eqnarray}
&&\int_0^\infty \kappa^{\nu+1}K_\mu(\alpha\kappa)I_\mu(\beta\kappa)J_\nu(\gamma\kappa)
={(\alpha\beta)^{-\nu-1}\gamma^\nu e^{-(\nu+1/2)\pi i}\over \sqrt{2\pi}(\Theta^2-1)^{\nu/2+1/4}}
{\cal D}_{\mu-1/2}^{\nu+1/2}(\Theta) ,
\end{eqnarray}
which holds for $\Re~\alpha>|\Re~\beta|$, $c>0$, $\Re~\nu>-1$, and $\Re~(\nu+\mu)>-1$, where 
$\Theta$ is defined as $2\alpha\beta\Theta=\alpha^2+\beta^2+\gamma^2$, and 
\begin{eqnarray}
{\cal D}_\nu^{1/2}(\Theta)=i\sqrt{\pi\over 2}(\Theta^2-1)^{-1/4}\left[\Theta+\sqrt{\Theta^2-1}\right]^{-\nu-1/2} ,
\end{eqnarray}
 we can show the following equality: 
\begin{eqnarray}
\int_0^\infty d\kappa \kappa J_{-i\omega/a}\left({\kappa\over a}e^{a\eta}\right)
K_{-i\omega/a}\left({\kappa\over a}\right) J_0(\kappa x_\perp)
={a\over 2\rho_0(x)}e^{i\omega\tau_+^{x}-i\omega\zeta}.
\end{eqnarray}
Here we used the definitions of $\rho_0(x)$ and $\tau_\pm^x$ in (\ref{defrho0}) and (\ref{deftaum}), respectively.

Then, (\ref{firsttermMppM}) 
reduces to 
\begin{eqnarray}
&&\int_0^{\infty}d\omega\int\int d^2k_\perp
v_{\omega,\bm k_\perp}^{\rm F,d}(x)v_{\omega,\bm k_\perp}^{\rm R}(z(\tau))
{e^{\pi\omega/a}\over e^{2\pi\omega/a}-1}
={-i\over 8\pi^2\rho_0(x)}\int_0^\infty d\omega e^{-i\omega\tau +i\omega\tau_+^{x}}
{e^{\pi\omega/a}\over e^{2\pi\omega/a}-1},
\label{firsttermMppMA}
\end{eqnarray}
and 
\begin{eqnarray}
&&\langle0,{\rm M}|\phi_{\rm h}^{\rm F,d}(x) \phi_{\rm h}(z(\tau))|0,{\rm M}\rangle
={-i\over 8\pi^2\rho_0(x)}\int_{-\infty}^{\infty} d\omega e^{-i\omega\tau}
{e^{\pi\omega/a}\over e^{2\pi\omega/a}-1}e^{i\omega\tau_+^{x}}.
\end{eqnarray}
This is nothing but the first term of Eq.~(\ref{resultoffs}) for $x$ in the F region.

Similarly, the second term of the integration on the right-hand side of (\ref{MppM2}) is 
evaluated as 
\begin{eqnarray}
&&\int_0^{\infty}d\omega\int\int d^2k_\perp
v_{\omega,\bm k_\perp}^{\rm F,s}(x)v_{\omega,\bm k_\perp}^{\rm R*}(z(\tau))
{1\over 1-e^{-2\pi\omega/a}}
\nonumber\\
&&~~={-i\over \sqrt{16 a^2\pi^4}}\int_0^{\infty}d\omega e^{-i\omega\zeta+i\omega\tau}
{1\over 1-e^{-2\pi\omega/a}} \int_0^\infty d\kappa \kappa J_{-i\omega/a}\left({\kappa\over a}e^{a\eta}\right)
K_{i\omega/a}\left({\kappa\over a}\right) J_0(\kappa x_\perp)
\nonumber\\
&&~~={-i\over 8\pi^2\rho_0(x)}\int_0^\infty d\omega e^{i\omega\tau -i\omega\tau_-^{x}}
{1\over 1-e^{-2\pi\omega/a}},
\label{firsttermMppMD}
\end{eqnarray}
where we used $K_\nu(z)=K_{-\nu}(z)$ and the relation
\begin{eqnarray}
{a\over 2\rho_0(x)}e^{i\omega\tau_+^{x}-i\omega\zeta}
={a\over 2\rho_0(x)}e^{-i\omega\tau_-^{x}+i\omega\zeta},
\label{arhoarho}
\end{eqnarray}
which is obtained directly from the definition of $\tau_\pm^x$ (see Appendix C).
Thus, we finally obtain
\begin{eqnarray}
&&\langle0,{\rm M}|\phi_{\rm h}^{\rm F,s}(x) \phi_{\rm h}(z(\tau))|0,{\rm M}\rangle
={-i\over 8\pi^2\rho_0(x)}\int_{-\infty}^{\infty} d\omega e^{-i\omega\tau}\left(
-{e^{i\omega\tau_-^{x}}\over e^{2\pi\omega/a}-1}
\right),
\end{eqnarray}
which is the second term in (\ref{resultoffs}).

The results in this section demonstrate that 
$\langle0,{\rm M}| \phi_{\rm h}^{\rm F,d}(x) \phi_{\rm h}(z(\tau))|0,{\rm M}\rangle$
and 
$\langle0,{\rm M}| \phi_{\rm h}^{\rm F,s}(x) \phi_{\rm h}(z(\tau))|0,{\rm M}\rangle$
explain the first and second terms in (\ref{resultoffs}), 
which correspond to the first and second terms in 
the interference term (\ref{interferenceall}), respectively. 
Therefore, the remaining two interference terms come from 
$\langle0,{\rm M}| \phi_{\rm h}^{\rm F,d}(x) \phi_{\rm h}(z(\tau))|0,{\rm M}\rangle$. 
As shown in the previous section, 
the right-moving wave Kasner mode and the Rindler mode, 
$v_j^{\rm F,d}(=v^{\rm II}_j)$ and $v_j^{\rm R}(=v^{\rm I}_j)$, respectively, are in the entangled state; 
therefore, this can be understand as the origin of the quantum radiation produced 
by the Unruh--DeWitt detector. 

\subsection{Physical interpretations}
In this section, we consider the physical meaning of the cancellation of the
naive radiation term by the interference terms. 
In the equilibrium phase, the Unruh--DeWitt detector is thermalized at the Unruh temperature, 
and the inhomogeneous solution, $\phi_{\rm inh}$, is given by the stationary solution. 
In this phase, one may expect that the total radiation will vanish because 
the outgoing and incoming fluxes are balanced.
However, the present situation differs from that in an ordinary thermalized system because 
the harmonic oscillator is accelerated, and energy is constantly injected. 
Furthermore, the thermal behavior of the system is obtained only by tracing out the
states in the L Rindler wedge, but in the F region, into which most of the radiation flux propagates, 
we cannot separate the L and R Rindler modes, and 
the logic based on the thermal behavior of the Unruh effect is not necessary valid. 
In the following, we discuss the origin of the partial cancellation
of the radiation flux in terms of the Kubo--Martin--Schwinger (KMS) relation.

As we saw in \cite{IYZ}, if the KMS-like relation
\begin{eqnarray}
&& \langle \varphi(\omega) \phi_{\rm h} (x) \rangle = \rho(\omega) \langle [\varphi(\omega), \phi_{\rm h}(x)] \rangle
\end{eqnarray}
is satisfied, the two-point correlation function is shown to
reduce to 
\begin{eqnarray}
&& \langle \phi(x) \phi(y) \rangle  - \langle \phi_{\rm h}(x) \phi_{\rm h}(y) \rangle = 
-\frac{ i \lambda^2 }{2 \pi^2} \int d \tau_x d \tau_y d \omega e^{-i \omega (\tau_x -\tau_y)} \rho(\omega)
\nonumber \\
&& \times \left[ G_R(x, z(\tau_x)) G_A(y,z(\tau_y)) f(\omega) h(\omega) 
-G_A(x, z(\tau_x)) G_R(y,z(\tau_y)) f(-\omega) h(-\omega) \right].
\end{eqnarray}
Here  $\rho(\omega)$ is any function of $\omega$ and is typically given by $\rho(\omega)=1/(1-e^{- \omega/T})$.

In the F region, because $x$ (or $y$) is always in the future of the trajectory $z(\tau)$, 
$G_A(x, z(\tau))$ vanishes. Thus, if the above relation holds, the two-point correlation would vanish. 
However, a straightforward calculation shows \cite{IYZ} that the relation is slightly violated as
\begin{eqnarray}
&& \langle \varphi(\omega) \phi_{\rm h} (x) \rangle = \rho(\omega) \langle [\varphi(\omega), \phi_{\rm h}(x)] \rangle
- \frac{i}{4\pi \rho_0(x)} \frac{e^{\pi \omega/a}}{e^{2\pi \omega/a}-1} e^{i \omega \tau_+(x)},
\end{eqnarray}
where $\rho(\omega)=1/(1-e^{-2\pi \omega/a})$. The second term, which violates the KMS-like relation, is responsible
for the quantum Unruh radiation.

\section{Application II: Quantum radiation from a uniformly accelerating detector in two-dimensional spacetime
\label{Sec:radiation-in-d2} }

As an application of the results in Sec.~IV, we study 
the Unruh--DeWitt detector model coupled to a massless scalar field in two-dimensional spacetime. 
We first review the derivation of the two-point function in this model, following Ref.~\cite{HuRaval}. 
The action is given by (\ref{actionUD}) with the dimensionality replaced by $d=2.$
The equations of motion for $Q(\tau)$ and $\phi(x)$ are essentially the same as 
(\ref{eqQ}) and (\ref{eqphiA}), respectively. 
Hereafter, we adopt $P[Q]={dQ/d\tau}$ and $\bar P[\phi(z(\tau))]=-{d\phi(z(\tau))/d\tau}$
for simplicity, as in Ref.~\cite{HuRaval}.
As in the previous section, the equations can be solved exactly, and 
the solution of $\phi$ is given as the sum of the homogeneous solution $\phi_{\rm h}(x)$ and the 
inhomogeneous solution $\phi_{\rm inh}(x)$, i.e.,
$  \phi(x)=\phi_{\rm h}(x)+\phi_{\rm inh}(x)$,
where $\phi_{\rm h}(x)$ carries the vacuum fluctuation, and $\phi_{\rm inh}(x)$ is given 
in terms of $Q(\tau)$ as
\begin{eqnarray}
\phi_{\rm inh}(x) =\lambda \int d\tau' {d\over d\tau'}Q(\tau')G_R(x-z(\tau')).
\label{phiQG2}
\end{eqnarray}
In the two-dimensional case, the retarded Green function for a massless scalar field is given by
\begin{eqnarray}
G_R(x-z(\tau'))={1\over 2}\theta(t-z+e^{a\tau'}/a)\theta(t+z-e^{a\tau'}/a).
\label{RGF2}
\end{eqnarray}
Thus, we find that the equation of motion for $Q(\tau)$ reduces to 
\begin{eqnarray}
&&\ddot Q(\tau)+2\gamma\dot Q(\tau)+\Omega_0^2Q(\tau)
=-{\lambda\over m}{d\over d\tau}\phi_{\rm h}(z(\tau)),
\label{eqQ2b}
\end{eqnarray}
where we defined $\gamma={\lambda^2/ 4m}$.
The solution is given by 
\begin{eqnarray}
&&Q(\tau)={-}{\lambda\over m}\int_{-\infty}^\tau d\tau' {e^{-\gamma(\tau-\tau')}\sin(\Omega_0(\tau-\tau'))\over \Omega}
{d\phi_{\rm h}(z(\tau'))\over d\tau'}. 
\end{eqnarray}
We find that the two-point function is given by 
\begin{eqnarray}
&&\langle\phi_{\rm }(x)\phi_{\rm }(x')\rangle-\langle\phi_{\rm h}(x)\phi_{\rm h}(x')\rangle
=
\langle\phi_{\rm h}(x)\phi_{\rm inh}(x')\rangle
+\langle\phi_{\rm inh}(x)\phi_{\rm h}(x')\rangle
+\langle\phi_{\rm inh}(x)\phi_{\rm inh}(x')\rangle
\nonumber\\
&&~~~~~~~~~~~~~~
=-{\gamma\over 2\pi}\int_{-\infty}^{\infty}{d\omega\over \omega}{e^{\pi\omega/a}\over e^{2\pi\omega/a}-1}
\biggl\{|a^2 UV'|^{-i\omega/a}h(\omega)+|a^2 U'V|^{i\omega/a}h(-\omega)
\biggr\}
\label{tdfinal}
\end{eqnarray}
when $x$ and $x'$ are in the F region Ref.~\cite{HuRaval}. 
The calculations are given in Sec.~\ref{Sec:2dimGreen}.
We defined $h(\omega)=i\omega/(\omega^2-\Omega_0^2+2i\omega\gamma)$. 
When $x$ and $x'$ are in the right-hand region of the detector's trajectory in 
the R region, i.e., $z^2-t^2>1/a^2$ and $z'{}^2-t'{}^2>1/a^2$, the two-point function is
\begin{eqnarray}
&&\langle\phi_{\rm }(x)\phi_{\rm }(x')\rangle-\langle\phi_{\rm h}(x)\phi_{\rm h}(x')\rangle
=
\langle\phi_{\rm h}(x)\phi_{\rm inh}(x')\rangle
+\langle\phi_{\rm inh}(x)\phi_{\rm h}(x')\rangle
+\langle\phi_{\rm inh}(x)\phi_{\rm inh}(x')\rangle
\nonumber\\
&&~~~~~~~~~~~~~~
=-{\gamma\over 2\pi}\int_{-\infty}^{\infty}{d\omega\over \omega}{1\over e^{2\pi\omega/a}-1}
\biggl\{|a^2 UV'|^{-i\omega/a}h(-\omega)+|a^2 U'V|^{i\omega/a}h(\omega)
\biggr\}.
\label{tdfinalR}
\end{eqnarray}
Note that the two-point function is a function of either $UV'$ or $U'V$.
The two-point function with the vacuum contribution subtracted is nonzero,
although
the quantum radiation vanishes, as shown in Ref.~\cite{HuRaval}.
This can be understood as follows. 
The radiation flux is calculated as
$T_{tz}=T_{zt}=T_{VV}-T_{UU}$ with $T_{UU}=\lim_{U'\rightarrow U} 
\partial_U\partial_{U'}
[\langle \phi(x)\phi(x')\rangle-\langle\phi_{\rm h}(x)\phi_{\rm h}(x')\rangle]$ 
and $T_{VV}=\lim_{V'\rightarrow V} 
\partial_V\partial_{V'}
[\langle \phi(x)\phi(x')\rangle-\langle\phi_{\rm h}(x)\phi_{\rm h}(x')\rangle]$,
but the two-point function has no dependence on $UU'$ or $VV'$. 
Therefore, there is no quantum radiation in the two-dimensional model.
This behavior is specific to the two-dimensional case. 

If the naive radiation term was not canceled by the interference term,
it would give a nonvanishing flux even in the two-dimensional case. 
Actually, when $x$ and $x'$ are in the F region, the naive radiation term is given by 
(Ref.~\cite{HuRaval})
\begin{eqnarray}
&&\langle\phi_{\rm inh}(x)\phi_{\rm inh}(x')\rangle
={2\gamma^2\over \pi}\int_{-\infty}^{\infty}{d\omega\over \omega}{1\over e^{2\pi\omega/a}-1}
(V/V')^{i\omega/a}|h(\omega)|^2.
\end{eqnarray}
Then if we use the relation $h(\omega)+h(-\omega)=4\gamma|h(\omega)|^2$, the 
the naive radiation term is rewritten as
\begin{eqnarray}
&&\langle\phi_{\rm inh}(x)\phi_{\rm inh}(x')\rangle
={\gamma\over 2\pi}\int_{-\infty}^{\infty}{d\omega\over \omega}{1\over e^{2\pi\omega/a}-1}
(V/V')^{i\omega/a}(h(\omega)+h(-\omega)).
\label{pxpxvv}
\end{eqnarray}
Thus, if this term remained, the component of the energy--momentum tensor $T_{VV}$ would be
nonzero. 
\subsection{Calculation of $\langle\phi_{\rm h}(x)\phi_{\rm h}(z(\tau))\rangle$ 
in the reference frame coordinates
\label{Sec:2dimGreen}}
In the following, we look at the details of the calculations of the interference terms
and compare the Green function method in the reference frame coordinates 
with the calculation in the operator formalism with (\ref{MIII}) 
to determine the origin of the two-point function (\ref{tdfinal}). 
In particular, we focus on the Wightman function $\langle \phi_{\rm h}(x)\phi_{\rm h}(z(\tau'))\rangle$
for $x$ in the F region. 

The two-dimensional massless scalar field is described by (\ref{phitz2}). 
According to Ref.~\cite{HuRaval}, the scalar field on the accelerated 
trajectory of the detector can be written as
\begin{eqnarray}
&&\phi_{\rm h}(z(\tau^{\prime}))
=
{1\over 2 \pi a}
\int_{-\infty}^{\infty} {dk\over \sqrt{4\pi |k|}} 
\biggl[\hat b_k\biggl\{
\int_{-\infty}^{\infty} d\omega 
e^{-i\omega \tau'} e^{\pi \omega / 2a}
\Gamma(-{i\omega/ a}) \left|{k/ a}\right|^{i\omega / a}\theta(k) 
\nonumber\\
&&~~~~~~~~~~~~~~~
+
\int_{-\infty}^{\infty} d\omega 
e^{-i\omega \tau'} e^{\pi \omega / 2a}
\Gamma({i\omega / a}) \left|{k / a}\right|^{-i\omega / a} \theta(-k) \biggr\}
+h.c.\biggr].
\label{2MphiO}
\end{eqnarray}
Then, for $x$ in the F region, we find the following formula:
\begin{eqnarray}
\langle 0,{\rm M}| 
\phi_{\rm h}(x) \phi_{\rm h} (z(\tau'))|0, {\rm M}\rangle
&=&{1\over {4 \pi }}\int_{-\infty}^{\infty} {d\omega\over \omega}
e^{i\omega \tau^{\prime}} 
{1\over e^{\pi\omega/a}-e^{-\pi\omega/a}}
\bigl\{
({aU})^{{i\omega / a}} 
+  e^{\pi\omega/a}({aV})^{-i\omega /a}\bigr\}
\nonumber \\
&=&{1\over {4 \pi }}\int_{-\infty}^{\infty} {d\omega\over \omega}
e^{-i\omega \tau^{\prime}} 
{1\over e^{2\pi\omega/a}-1}
\bigl\{e^{\pi\omega/a}
({aU})^{{-i\omega / a}} 
+  ({aV})^{i\omega /a}\bigr\},
\label{MMM}
\end{eqnarray}
where we used  
\begin{eqnarray}
&&\int_0 ^{\infty} {dk \over |k|} e^{-ikU}
\left|{k \over a}\right|^{-i\omega / a}
=
e^{-{\pi \omega / 2a}}
\bigl( { aU} \bigr) ^{{i\omega /a}}
\Gamma(-{i\omega /a}),
\nonumber\\
&&\int_{-\infty} ^0 {dk \over |k|} e^{ikV}
\left|{k \over a}\right|^{i\omega / a} 
=
e^{\pi \omega / 2a}
\bigl( {aV} \bigr) ^{-i\omega /a}
\Gamma({i\omega/ a}).
\nonumber
\end{eqnarray}
The first and second terms on the right-hand side of (\ref{MMM})
correspond to the first and second terms on the right-hand side of 
(\ref{tdfinalR}).
The second term cancels the naive radiation term, as in the four-dimensional case.
Thus, the first term is responsible for the remaining two-point function (\ref{tdfinal}).
In the next subsection, we perform the same calculation
using the operator formalism with (\ref{MIIItwo}) to determine the physical origin of 
these two terms in (\ref{MMM}).

\subsection{Calculation of $\langle\phi_{\rm h}(x)\phi_{\rm h}(z(\tau))\rangle$ in the operator formalism with 
(\ref{MIIItwo})}
Here, the same calculation is performed in the operator formalism with (\ref{MIIItwo}).
Following the description in Sec.~\ref{Sec:entangeD=2}, 
when $x$ is in the F region, one can write the quantized field 
in two-dimensional spacetime as 
\begin{eqnarray}
\phi_{\rm h}(x)=\phi_{\rm h}^{\rm F,d}(x)+\phi_{\rm h}^{\rm F,s}(x),
\end{eqnarray}
where we defined
\begin{eqnarray}
&&\phi_{\rm h}^{\rm F,d}(x)=\int_{0}^{\infty} {dk\over \sqrt{4\pi k}}  \left(\hat a^{\rm III}_\omega
v_\omega^{\rm III}(x) +h.c.\right),
\\
&&\phi_{\rm h}^{\rm F,s}(x)=\int_{0}^{\infty} {dk\over \sqrt{4\pi k}}  
\left(\hat a^{\rm IV}_\omega
v_\omega^{\rm IV}(x) +h.c.\right).
\end{eqnarray}
The quantum field on the detector's trajectory is given by 
\begin{eqnarray}
\phi_{\rm h}(z(\tau'))=\phi_{\rm h}^{\rm R,d}(z(\tau'))+\phi_{\rm h}^{\rm R,s}(z(\tau')),
\end{eqnarray}
where we defined
\begin{eqnarray}
&&\phi_{\rm h}^{\rm R,d}(z(\tau'))=\int_{0}^{\infty} {dk\over \sqrt{4\pi k}}  
 \left(\hat a^{\rm I}_\omega
v_\omega^{\rm I}(z(\tau')) +h.c.\right),
\\
&&\phi_{\rm h}^{\rm R,s}(z(\tau'))=\int_{0}^{\infty} {dk\over \sqrt{4\pi k}}  
 \left(\hat a^{\rm IV}_\omega
v_\omega^{\rm IV}(z(\tau')) +h.c.\right).
\end{eqnarray}
Thus, the two-point function is given in the following form:
\begin{eqnarray}
\langle 0,{\rm M}| \phi_{\rm h}(x) \phi_{\rm h} (z(\tau'))|0, {\rm M}\rangle
&=&
\langle 0,{\rm M}| \phi_{\rm h}^{\rm F,d}(x) \phi_{\rm h}^{\rm R,d}(z(\tau'))|0, {\rm M}\rangle
+\langle 0,{\rm M}| \phi_{\rm h}^{\rm F,s}(x) \phi_{\rm h}^{\rm R,s}(z(\tau'))|0, {\rm M}\rangle,
\end{eqnarray}
where we defined
\begin{eqnarray}
&&\langle 0,{\rm M}| \phi_{\rm h}^{\rm F,d}(x) \phi_{\rm h}^{\rm R,d}(z(\tau'))|0, {\rm M}\rangle
=
{1\over 4\pi}\int_{-\infty}^\infty {d\omega\over \omega}
{e^{\pi\omega/a}\over e^{2\pi\omega/a}-1}
(aU)^{-{{i \omega}/ a}}e^{-i\omega\tau'},
\\
&&\langle 0,{\rm M}| \phi_{\rm h}^{\rm F,s}(x) \phi_{\rm h}^{\rm R,s}(z(\tau'))|0, {\rm M}\rangle
=
{1\over 4\pi}\int_{-\infty}^\infty {d\omega\over \omega}
{1\over e^{2\pi\omega/a} -1}
(aV)^{{i \omega}/ a}
e^{-i\omega\tau'}.
\end{eqnarray}
Here we used the relations derived in Sec.~\ref{Sec:entangeD=2}. 
Note that these two formulas are equal to the first 
and second terms on the right-hand side of (\ref{MMM}), respectively. 
The detector's trajectory is parameterized by 
$x'={a}^{-1}\cosh a\tau'$ and $t'={a}^{-1}\sinh a\tau'$, 
and we can write $e^{a\tau^{\prime}}=a(x^{\prime}+t^{\prime})=aV^{\prime}$
on the detector' trajectory. 
Then, we may write the two-point Wightman function as 
\begin{eqnarray}
&&\langle 0,{\rm M}| \phi_{\rm h}^{\rm F,d}(x) \phi_{\rm h}^{\rm R,d}(z(\tau'))|0, {\rm M}\rangle
=
{1\over 4\pi}\int_{-\infty}^\infty {d\omega\over \omega}
{e^{\pi\omega/a}\over e^{2\pi\omega/a}-1}
(a^2 UV^{\prime} )^{-i\omega/ a},
\label{MdfrdM}
\\
&&\langle 0,{\rm M}| \phi_{\rm h}^{\rm F,s}(x) \phi_{\rm h}^{\rm R,s}(z(\tau'))|0, {\rm M}\rangle
=
{1\over 4\pi}\int_{-\infty}^\infty {d\omega\over \omega}
{1\over e^{2\pi\omega/a} -1}
(V /V^{\prime})^{i\omega / a}.
\label{MsfsdM}
\end{eqnarray}
As in the four-dimensional case studied in the previous section, (\ref{MdfrdM}), 
$\langle 0,{\rm M}| \phi_{\rm h}^{\rm F,d}(x) \phi_{\rm h}^{\rm R,d}(z(\tau'))|0, {\rm M}\rangle$
is almost the same as the first term on the right-hand side of (\ref{tdfinal}). 
The differences between them are the coupling constant and $h(\omega)$, which reflects the
interaction between the detector and the vacuum fluctuations. 
Therefore, the remaining interference terms come from 
$\langle 0,{\rm M}| \phi_{\rm h}^{\rm F,d}(x) \phi_{\rm h}^{\rm R,d}(z(\tau'))|0, {\rm M}\rangle$, 
whereas the contribution corresponding to 
$\langle 0,{\rm M}| \phi_{\rm h}^{\rm F,s}(x) \phi_{\rm h}^{\rm R,s}(z(\tau'))|0, {\rm M}\rangle$ 
is canceled out by the naive radiation term (\ref{pxpxvv}). 

In contrast to the four-dimensional case, as described in Ref.~\cite{HuRaval}, 
the remaining two-point function does not give rise to quantum radiation in 
the two-dimensional case. 
This property is a special characteristics of the two-dimensional case. 

\section{Summary and Conclusions}
In this paper, we extended the entanglement structure of the Minkowski vacuum 
from the ordinary L and R Rindler wedges into the entire 
Minkowski spacetime, including 
the Kasner expanding universe (F region) and Kasner shrinking universe (P region). 
Our result clarifies the structure of the entanglement of the Minkowski vacuum state in a unified manner and 
makes it possible to give an operator interpretation of the calculation of
the two-point correlation functions.

We also applied the results to discuss the physical origin of 
 the quantum radiation produced by an Unruh--DeWitt detector
in uniformly accelerated motion. 
We showed that quantum entanglement between the Rindler mode 
in the R region and the right-moving wave Kasner mode in the F region explains the origin of 
 the quantum radiation. 
In the two-dimensional case, a similar structure appears in the two-point function
of the field; however, the energy--momentum tensor vanishes owing to a special characteristic
of two-dimensional spacetime.

\vspace{2mm}
{\it Acknowledgments.}---
This work is supported by MEXT/JSPS KAKENHI Grant Numbers 15H05895, 
17K05444, and 17H06359 (KY) and 23540329 and 16H06490 (SI). 
We thank R.~Tatsukawa, Y.~Nambu, J.~Soda, S.~Ishizaka, M.~Iinuma, H.~F.~Hofmann,
M. Hotta, R.~Schutzhold, W.~G.~Unruh, P.~Chen, S-Y.~Lin, and B.~L.~Hu for useful comments.
K.Y. is grateful for the warm hospitality of the Mathematics Department 
of the University of York, where we initiated this work.

\appendix

\section{Analytic continuation of the mode functions}
In this appendix, we show that the 
 relations of the mode functions derived in Sec.~III can also be obtained by using the continuation 
through the Minkowski positive-frequency mode function.  
Here we first compute the purely positive-frequency mode in the F and P regions. 
Following (\ref{definitionw}), 
we have 
\begin{eqnarray}
w_{\pm \omega,{\bm k}_\perp}&=&{e^{i{\bm k}_\perp\cdot{\bm x}_\perp}\over \sqrt{8a}2\pi^2} e^{\pm i\zeta\omega}
\int_{-\infty}^{\infty} d\theta e^{\pm i\theta\omega/a}
e^{-i(\kappa e^{a\eta}/a)\cosh\theta }
\nonumber\\
&=&-i{e^{i{\bm k}_\perp\cdot{\bm x}_\perp}\over \sqrt{8a}2\pi} e^{\pm i\zeta\omega}e^{\pi\omega/2a}
H_{i\omega/a}^{(2)}(\kappa e^{a\eta}/a)
\label{wF}
\end{eqnarray}
for the F region and 
\begin{eqnarray}
w_{\pm \omega,{\bm k}_\perp}&=&{e^{i{\bm k}_\perp\cdot{\bm x}_\perp}\over \sqrt{8a}2\pi^2} e^{\mp i\tilde\zeta\omega}
\int_{-\infty}^{\infty} d\theta e^{\pm i\theta\omega/a}
e^{+i (\kappa e^{-a\widetilde\eta}/a)\cosh\theta }
\nonumber\\
&=&+i{e^{i{\bm k}_\perp\cdot{\bm x}_\perp}\over \sqrt{8a}2\pi} e^{\mp i\tilde\zeta\omega}e^{-\pi\omega/2a}
H_{i\omega/a}^{(1)}(\kappa e^{-a\tilde\eta}/a)
\label{wP}
\end{eqnarray}
for the P region, 
where we used
the formulas for the Hankel function
\begin{eqnarray}
&&H^{(1)}_\nu(z)={-2ie^{-\nu\pi i/2}\over \pi}\int_0^\infty e^{+iz\cosh t}\cosh \nu t dt,
\\
&&H^{(2)}_\nu(z)={+2ie^{+\nu\pi i/2}\over \pi}\int_0^\infty e^{-iz\cosh t}\cosh \nu t dt.
\end{eqnarray}
Note that all the positive-frequency mode functions can be constructed within the F or P region,
even for the massless field \cite{Higuchi,Fulling,BD,Padmanabhan}. 
This will be true for a $d(\geq 3)$-dimensional
spacetime because the transverse momentum acts as an effective mass. 
When the field has a mass, all the information goes to the F region or comes from the P region.
Hence, one can construct the mode function corresponding to the Minkowski positive-frequency mode
in the F and P regions.

\subsection{Continuation to the positive-frequency Minkowski mode function}
Eqs.~(\ref{wF}) and (\ref{wP}) can be analytically continued as a 
Minkowski positive-frequency solution to the R and
L Rindler wedges as follows. We first note that, because positive-frequency solutions behave 
like $e^{-ik_0 t}$, it is implicit that upon analytic continuation,
we must treat $t$ as $t-i\epsilon$, where $\epsilon > 0$, so that the exponential function does 
not diverge as $k_0\to\infty$. On the other hand, for negative-frequency solutions behaving 
like $e^{ik_0 t}$, it is implicit that we must treat $t$ as $t+i\epsilon$.

Now, letting $t\pm z \to t \pm z -i\epsilon$, as $t\pm z$ changes from a positive to a negative value, 
we have a small imaginary part. This means that
$(t - z)^b \to e^{-i\pi b}(z-t)^b$ as $t-z$ becomes negative (R Rindler wedge) and
$(t+z)^b \to e^{-i\pi b}(-z-t)^b$ as $t+z$ becomes negative (L Rindler wedge).  
Table \ref{tableone} summarizes the continuation of $\sqrt{t^2-z^2}$ and 
$\sqrt{(t+z)/(t-z)}$ from the F and P regions to the R and L regions through the horizons.
These continuation rules are equivalent to the continuation of the coordinate variables 
shown in Table II.

\begin{center}
\begin{table}[t]
\begin{tabular}{lll}
\hline\\
~~~~~~
 ${\rm F} \longrightarrow {\rm R}$ ~~~&
 $\displaystyle{\sqrt{t^2-z^2}\rightarrow e^{-\pi i/2}\sqrt{z^2-t^2}}$, ~~~~&
 $\displaystyle{\sqrt{t+z\over t-z}\rightarrow e^{+\pi i/2} \sqrt{z+t\over z-t}}$
~~~~~~
\\
\\
~~~~~~
 ${\rm F} \longrightarrow {\rm L}$ ~~~&
 $\displaystyle{\sqrt{t^2-z^2}\rightarrow e^{-\pi i/2}\sqrt{z^2-t^2}}$, ~~~~&
 $\displaystyle{\sqrt{t+z\over t-z}\rightarrow e^{-\pi i/2} \sqrt{z+t\over z-t}}$
\\
\\
~~~~~~
 ${\rm P}  \longrightarrow {\rm R}$ ~~~&
 $\displaystyle{\sqrt{t^2-z^2}\rightarrow e^{+\pi i/2}\sqrt{z^2-t^2}}$, ~~~~&
 $\displaystyle{\sqrt{t+z\over t-z}\rightarrow e^{+\pi i/2} \sqrt{z+t\over z-t}}$
\\
\\
~~~~~~
 $ {\rm P}  \longrightarrow {\rm L}$ ~~~&
 $\displaystyle{\sqrt{t^2-z^2}\rightarrow e^{+\pi i/2}\sqrt{z^2-t^2}}$, ~~~~&
 $\displaystyle{\sqrt{t+z\over t-z}\rightarrow e^{-\pi i/2} \sqrt{z+t\over z-t}}$
\\
\\
\hline
\end{tabular}
\caption{Continuation of $\sqrt{t^2-z^2}$ and $\sqrt{(t+z)/(t-z)}$ 
from F and P regions to R and L regions.}
\label{tableone}
\vspace{1cm}
\begin{tabular}{lll}
\hline\\
~~~~~~
 ${\rm F} \longrightarrow {\rm R}$ ~~~&
 $\displaystyle{\tau=\zeta-{\pi \over 2a }i}$, ~~~~&
 $\displaystyle{\xi=\eta+{\pi \over 2a}i}$~~~~~~
\\
\\
~~~~~~
 ${\rm F} \longrightarrow {\rm L}$ ~~~&
 $\displaystyle{\tilde \tau =-\zeta-{\pi \over 2a }i}$,~~~~&
 $\displaystyle{\tilde \xi = \eta+{\pi \over 2a}i}$
\\
\\
~~~~~~
 ${\rm P}  \longrightarrow {\rm R}$ ~~~&
 $\displaystyle{\tau=-\tilde\zeta-{\pi \over 2a }i}$,~~~~&
 $\displaystyle{\xi=-\tilde\eta-{\pi \over 2a}i}$
\\
\\
~~~~~~
 $ {\rm P}  \longrightarrow {\rm L}$ ~~~&
 $\displaystyle{\tilde\tau=\tilde\zeta-{\pi \over 2a }i}$,~~~~&
 $\displaystyle{\tilde\xi=-\tilde\eta-{\pi \over 2a}i}$\\
\\
\hline
\end{tabular}
\label{tabletwo}
\caption{Continuation of the coordinate variables from F and P regions to R and L regions.}
\end{table}
\end{center}

\subsection{Continuation from F region to R and L regions}
From Sec.~II.C, we choose the positive-frequency mode function in the F region: 
\begin{eqnarray}
&&\hspace{-5mm}
v_{\omega,{\bm k}_\perp}^{\F,\s}(x)=v_{-\omega,{\bm k}_\perp}^\F(x)=
{-ie^{-i\omega\zeta}\over 2\pi\sqrt{4a\sinh (\pi \omega/a)}}
J_{-i\omega/a}\left({\kappa e^{a\eta}\over a}\right)
e^{i\bm k_\perp\cdot \bm x_\perp},
\label{vfs}
\\  &&\hspace{-5mm}
  v_{\omega,{\bm k}_\perp}^{\F,\d}(x)=v_{\omega,-{\bm k}_\perp}^\F(x)=
{-ie^{i\omega\zeta}\over 2\pi\sqrt{4a\sinh (\pi \omega/a)}}
J_{-i\omega/a}\left({\kappa e^{a\eta}\over a}\right)
e^{-i\bm k_\perp\cdot \bm x_\perp}.
  ~~
\label{vfd}
\end{eqnarray}
Using the mathematical formulas $J_{-\nu}(z)=[e^{-\nu\pi i}H_\nu^{(2)}(z)+e^{\nu\pi i}H_\nu^{(1)}(z)]/2$, 
$H_\nu^{(1)}(z)=(H_{\nu^*}^{(2)}(z^*))^*$, and $H_{-\nu}^{(2)}(z)=e^{-\nu\pi i}H_\nu^{(2)}(z)$, we have
\begin{eqnarray}
v^{\rm F,s}_{\omega,\bm k_\perp}&=& 
{-ie^{i\bm k_\perp\cdot \bm x_\perp}\over 2\pi\sqrt{4a\sinh (\pi \omega/a)}}
{1\over 2}
\left[e^{\pi \omega/a} e^{-i\omega\zeta} H_{i\omega/a}^{(2)}\left({\kappa e^{a\eta}\over a}\right)
+\left\{e^{i\omega\zeta} H_{i\omega/a}^{(2)}\left({\kappa e^{a\eta}\over a}\right)\right\}^*
\right]
\nonumber\\
&=&
{w_{-\omega,\bm k_\perp}-e^{-\pi\omega/a}w_{\omega,-\bm k_\perp}^*\over \sqrt{1-e^{-2\pi\omega/a}}},
\end{eqnarray}
\begin{eqnarray}
v^{\rm F,d}_{\omega,\bm k_\perp}&=& 
{-ie^{-i\bm k_\perp\cdot \bm x_\perp}\over 2\pi\sqrt{4a\sinh (\pi \omega/a)}}
{1\over 2}
\left[e^{\pi \omega/a} e^{i\omega\zeta} H_{i\omega/a}^{(2)}\left({\kappa e^{a\eta}\over a}\right)
+\left\{e^{-i\omega\zeta} H_{i\omega/a}^{(2)}\left({\kappa e^{a\eta}\over a}\right)\right\}^*
\right]
\nonumber\\
&=&
{w_{\omega,-\bm k_\perp}-e^{-\pi\omega/a}w_{-\omega,\bm k_\perp}^*\over \sqrt{1-e^{-2\pi\omega/a}}}.
\end{eqnarray}

Applying the continuation of the positive-frequency mode in the F region into the R region, 
$\displaystyle{\zeta\rightarrow\tau+{\pi \over 2a }i}$, 
 $\displaystyle{\eta\rightarrow\xi-{\pi \over 2a}i}$,
we have
\begin{eqnarray}
v^{\rm F,s}_{\omega,\bm k_\perp}&\rightarrow&{e^{i\bm x_\perp\cdot\bm k_\perp}}\sqrt{\sinh\pi\omega/a\over 4\pi^4a}e^{-i\omega\tau}K_{i\omega/a} 
\left({\kappa e^{a\xi}\over a}\right)=v^{\rm R}_{\omega,\bm k_\perp},
\\
v^{\rm F,d}_{\omega,\bm k_\perp}&\rightarrow&0 ,
\end{eqnarray}
where we used $K_\nu(z)=-(\pi i/2)e^{-\nu\pi i/2}H_\nu^{(2)}(e^{-\pi i/2}z)$ and $K_\nu(z)=K_{-\nu}(z)$.

Similarly, for the continuation of the positive-frequency mode in the F region into the L region, 
 $\displaystyle{ \zeta\rightarrow -\tilde\tau-{\pi \over 2a }i}$,
 $\displaystyle{ \eta\rightarrow\tilde\xi-{\pi \over 2a}i}$,
we have
\begin{eqnarray}
v^{\rm F,s}_{\omega,\bm k_\perp}&\rightarrow&0,
\\
v^{\rm F,d}_{\omega,\bm k_\perp}&\rightarrow&{e^{-i\bm x_\perp\cdot\bm k_\perp}}\sqrt{\sinh\pi\omega/a\over 4\pi^4a}e^{-i\omega\tilde\tau}K_{i\omega/a} 
\left({\kappa e^{a\tilde\xi}\over a}\right)=v^{\rm L}_{\omega,\bm k_\perp} .
\end{eqnarray}
This is the result described in Sec. III~F (Ref.~\cite{Higuchi}).

\subsection{Continuation from P region to R and L regions}
Using the relation $J_\nu(z)=(H_\nu^{(1)}(z)+H_\nu^{(2)}(z))/2$, 
$H_\nu^{(2)}(z)=(H_{\nu^*}^{(1)}(z^*))^*$
$H_{-\nu}^{(1)}(z)=e^{\nu\pi i}H_\nu^{(1)}(z)$,
the mode function in the P region is written as
\begin{eqnarray}
\hspace{-5mm}
v_{\omega,{\bm k}_\perp}^{\P,\s}(x)&=&{ie^{-i\bm k_\perp\cdot \bm x_\perp}
\over 2\pi\sqrt{4a\sinh (\pi \omega/a)}}
e^{-i\omega \tilde\zeta}J_{i\omega/a}\left({\kappa e^{-a\tilde\eta}\over a}\right)
\nonumber\\
&=&{ie^{-i\bm k_\perp\cdot \bm x_\perp}
\over 2\pi\sqrt{4a\sinh (\pi \omega/a)}}
{1\over 2}\left[
e^{-i\omega \tilde\zeta}H_{i\omega/a}^{(1)}\left({\kappa e^{-a\tilde\eta}\over a}\right)
+e^{-\pi\omega/a}\left\{e^{i\omega \tilde\zeta}H_{i\omega/a}^{(1)}\left({\kappa e^{-a\tilde\eta}\over a}\right)
\right\}^*
\right]
\nonumber\\
&=&
{w_{\omega,-\bm k_\perp}-e^{-\pi\omega/a}w_{-\omega,\bm k_\perp}^*\over \sqrt{1-e^{-2\pi\omega/a}}},
\label{vps}
\\
  \hspace{-5mm}
  v_{\omega,{\bm k}_\perp}^{\P,\d}(x)&=&
{ie^{i\bm k_\perp\cdot \bm x_\perp}\over 2\pi\sqrt{4a\sinh (\pi \omega/a)}}
e^{i\omega \tilde\zeta}
J_{i\omega/a}\left({\kappa e^{-a\tilde\eta}\over a}\right)
\nonumber\\
&=&{ie^{i\bm k_\perp\cdot \bm x_\perp}\over 2\pi\sqrt{4a\sinh (\pi \omega/a)}}
{1\over 2}\left[
e^{i\omega \tilde\zeta}H_{i\omega/a}^{(1)}\left({\kappa e^{-a\tilde\eta}\over a}\right)
+e^{-\pi\omega/a}\left\{e^{-i\omega \tilde\zeta}H_{i\omega/a}^{(1)}\left({\kappa e^{-a\tilde\eta}\over a}\right)
\right\}^*
\right]
\nonumber\\
&=&
{w_{-\omega,\bm k_\perp}-e^{-\pi\omega/a}w_{\omega,-\bm k_\perp}^*\over \sqrt{1-e^{-2\pi\omega/a}}}.
\label{vpd}
\end{eqnarray}

Applying the continuation of the positive-frequency mode in the P region into the R region, 
 $\displaystyle{\tilde\zeta\rightarrow-\tau-{\pi \over 2a }i}$,
 $\displaystyle{\tilde\eta\rightarrow-\xi-{\pi \over 2a}i}$,
we have
\begin{eqnarray}
v^{\P,s}_{\omega,\bm k_\perp}&\rightarrow&0,
\\
v^{\P,d}_{\omega,\bm k_\perp}&\rightarrow&{e^{i\bm x_\perp\cdot\bm k_\perp}}\sqrt{\sinh\pi\omega/a\over 4\pi^4a}e^{-i\omega\tau}K_{i\omega/a} 
\left({\kappa e^{a\xi}\over a}\right)=v^{\rm R}_{\omega,\bm k_\perp},
\end{eqnarray}
where we used $K_\nu(z)=(\pi i/2)e^{\nu\pi i/2}H_\nu^{(1)}(e^{\pi i/2}z)$.

Similarly, for the continuation of the positive-frequency mode in the P region into the L region, 
 $\displaystyle{\tilde\zeta\rightarrow\tilde\tau+{\pi \over 2a }i}$,
 $\displaystyle{\tilde\eta\rightarrow-\tilde\xi-{\pi \over 2a}i}$,
we have
\begin{eqnarray}
v^{\P,s}_{\omega,\bm k_\perp}&\rightarrow&{e^{-i\bm x_\perp\cdot\bm k_\perp}}\sqrt{\sinh\pi\omega/a\over 4\pi^4a}e^{-i\omega\tilde\tau}K_{i\omega/a} 
\left({\kappa e^{a\tilde\xi}\over a}\right)=v^{\rm L}_{\omega,\bm k_\perp},
\\
v^{\P,d}_{\omega,\bm k_\perp}&\rightarrow&0.
\end{eqnarray}

\section{Calculation of $\langle\phi_{\rm h}(x)\phi_{\rm h}(z(\tau))\rangle$ with Rindler state for $x$
in the R region}
Here, we consider the correlation function with $x$ in the R region using the description 
in the R Rindler spacetime. In this case, we have 
\begin{eqnarray}
&&\hspace{-1cm}
\langle0,M|\phi_{\rm h}(x)\phi_{\rm h}(z(\tau'))|0,M\rangle
\nonumber\\
&=&
\int_0^\infty d\omega \int \int d^2 k_\perp
\left[v_j^R(x)v_j^{\rm R*}(z(\tau')){1\over 1-e^{-2\pi \omega/a}}
+v_j^{\rm R*}(x)v_j^{\rm R}(z(\tau')){1\over e^{2\pi \omega/a}-1}\right],
\end{eqnarray}
where $x$ is specified by the coordinates $(\tau,\xi,\bm x_\perp)$.
We consider the first term of the integration, 
\begin{eqnarray}
&&\hspace{-1cm}
\int_0^\infty d\omega \int d^2k_\perp v_j^R(x)v_j^{\rm R*}(z(\tau'))
=\int_0^\infty d\omega {\sinh \pi \omega /a\over 2\pi a} e^{-i\omega(\tau-\tau')} 
\int_0^\infty d\kappa\kappa
K_{i\omega/a}(\kappa e^{a\xi}/a) K_{i\omega/a}(\kappa /a) J_0(\kappa x_\perp),
\label{vv}
\end{eqnarray}
where we used $K_\nu(x)=K_{-\nu}(x)$ and $\int_0^{2\pi}d\varphi e^{i\kappa x_\perp\cos\varphi}
=2\pi J_0(\kappa x_\perp)$. 
Using the mathematical 
formula \cite{typo}, we have
\begin{eqnarray}
&&\int_{0}^\infty d\kappa\kappa^{\nu+1}K_\mu(\alpha\kappa)K_\mu(\beta\kappa)J_\nu(\gamma \kappa) d\kappa
={1\over 2}\sqrt{\pi\over 2}{\gamma^\nu\over (\alpha\beta)^{\nu+1}}
\Gamma\left({\nu+\mu+1}\right)\Gamma\left({\nu-\mu+1}\right)(\Theta^2-1)^{-\nu/2-1/4}{\cal B}_{\mu-1/2}^{-\nu-1/2}(\Theta),
\nonumber\\
\end{eqnarray}
where $\Theta$ is defined as $2\alpha\beta\Theta={\alpha^2+\beta^2+\gamma^2}$, and 
${\cal B}_{\mu-1/2}^{-\nu-1/2}(\zeta)$ is the Legendre function, 
which is satisfied under the conditions Re$(\nu\pm\mu)>-1$ and Re$\nu>-1$, 
and the formula on page 172 of Ref. \cite{Magnus}, 
\begin{eqnarray}
{\cal B}_{i\omega/a-1/2}^{-1/2}(\Theta)={1\over \sqrt{2\pi}}{1\over i\omega/a}{1\over (\Theta^2-1)^{1/4}}
\left\{\left(\Theta+\sqrt{\Theta^2-1}\right)^{i\omega/a}-\left(\Theta+\sqrt{\Theta^2-1}\right)^{-i\omega/a}\right\}.
\end{eqnarray}
Then we have
\begin{eqnarray}
&&\hspace{-1cm}
\int_0^\infty d\kappa\kappa
K_{i\omega/a}(\kappa e^{a\xi}/a) K_{i\omega/a}(\kappa/a) J_0(\kappa x_\perp) 
\nonumber\\
&=&
{ \pi a^2 e^{-a\xi} 
 \over 4i\sinh (\pi\omega/a)} {1\over \sqrt {\Theta^2-1}}
\left\{\left(\Theta+\sqrt{\Theta^2-1}\right)^{i\omega/a}-\left(\Theta+\sqrt{\Theta^2-1}\right)^{-i\omega/a}\right\},
\label{KKJ}
\end{eqnarray}
where $\Theta$ has the expression $\Theta={(e^{a\xi}+e^{-a\xi}(1+a^2 x_\perp^2))/ 2}$. 
In the R region, we have
\begin{eqnarray}
&&{e^{a\xi}\sqrt{\Theta^2-1}\over a}= \rho_0(x),
\\
&&e^{a\tau}(\Theta+\sqrt{\Theta^2-1})^{\pm1} =e^{a\tau_\pm^x}.
\end{eqnarray}
Then, (\ref{vv}) reduces to
\begin{eqnarray}
\int_0^\infty d\omega \int d^2k_\perp v_j^R(x)v_j^{\rm R*}(z(\tau'))
&=&{ -i\over 8\pi^2\rho_0(x)} 
\int_0^\infty d\omega e^{i\omega\tau'} (e^{-i\omega\tau_-^x}-e^{-i\omega\tau_+^x}),
\end{eqnarray}
and finally we have 
\begin{eqnarray}
\langle0,M|\phi_{\rm h}(x)\phi_{\rm h}(z(\tau))|0,M\rangle
&=&{ -i\over 8\pi^2\rho_0(x)} 
\int_{-\infty}^{\infty} d\omega e^{-i\omega\tau} (e^{i\omega\tau_+^x}-e^{i\omega\tau_-^x}){1\over e^{2\pi\omega/a}-1}
\end{eqnarray}
for $x$ in the R region. This is equivalent to (\ref{resultoffs}) for $x$ in the R region.

\section{Derivation of (\ref{arhoarho})}
Using the definition of $\tau_-$, (\ref{deftaum}), we can write
\begin{eqnarray}
e^{a\tau_-}={a\over 2(t-z)}\left(-L^2+\sqrt{L^4+{4\over a^2}(t^2-z^2)}\right)
={2(t+z)\over a}{1\over L^2+\sqrt{L^4+{4\over a^2}(t^2-z^2)}}.
\end{eqnarray}
Using the definition of $\tau_+$ for $x^\mu$ in the F region, 
(\ref{deftaup}), we can write
\begin{eqnarray}
e^{a\tau_-}={t+z\over t-z}e^{-a\tau_+}.
\end{eqnarray}
For $x^\mu$ in the F region, $(t+z)/(t-z)=e^{2\eta\zeta}$; then, we have
\begin{eqnarray}
e^{a\tau_--a\zeta}=e^{-a\tau_++a\zeta}.
\end{eqnarray}


\end{widetext}

\begin{thebibliography}{99}
\bibitem{typo}
There seem to be typos in the last second formula on p. 103 of Ref. \cite{Magnus} and the first formula 
on p. 719 of Ref. \cite{Tables}.
\end{thebibliography}
\end{document}